\documentclass[prb,aps,twocolumn,showpacs,superscriptaddress]{revtex4}
\usepackage[dvips]{graphicx}

\usepackage{latexsym}
\usepackage{amsmath}
\usepackage{subfigure}
\usepackage{epsfig}
\usepackage{bm}
\usepackage{feyn}
\newcommand{\beq}{\begin{equation}}
\newcommand{\eeq}{\end{equation}}
\newcommand{\bea}{\begin{eqnarray}}
\newcommand{\eea}{\end{eqnarray}}
\newcommand{\kslash}{\not \! k}
\newcommand{\qslash}{\not \! q}
\newcommand{\pslash}{\not \! p}
\newcommand{\br}{\boldsymbol{r}}

\newcommand{\bx}{\boldsymbol{x}}
\newcommand{\by}{\boldsymbol{y}}

\begin{document}

\title{Neel order, quantum spin liquids and quantum criticality in two dimensions}

\author{Pouyan Ghaemi}
\affiliation{ Department of Physics, Massachusetts Institute of
Technology, Cambridge, Massachusetts 02139}

\author{T. Senthil}
\affiliation{ Department of Physics, Massachusetts Institute of
Technology, Cambridge, Massachusetts 02139}
\affiliation{Center for
Condensed Matter Theory, Indian Institute of Science, Bangalore,
India 560012}

\begin{abstract}
This paper is concerned with the possibility of a direct second order transition out of a collinear Neel phase
to a paramagnetic spin liquid in two dimensional quantum antiferromagnets. Contrary to conventional wisdom,
we show that such second order quantum transitions can potentially occur to certain spin liquid states
popular in theories of the cuprates. We provide a theory of this transition and study its universal
properties in an $\epsilon$ expansion. The existence of such a transition has a number of interesting
implications for spin liquid based approaches to the underdoped cuprates. In particular it
considerably clarifies
existing ideas for incorporating antiferromagnetic long range order into such a spin liquid based approach.

\end{abstract}

\maketitle

\section{Introduction}
In recent years a number of old ideas on theories of the cuprate materials have been clarified and
rejuvenated. A variety of experiments now support the notion that the underdoped cuprates are usefully
regarded as doped Mott insulators - in other words proximity to the Mott insulator
strongly influences the properties of the
underdoped materials\cite{OrMil}. The undoped Mott insulating parent materials have long range antiferromagnetic order.
However this order disappears rather quickly upon doping. Rather (at not too low temperatures) the pseudogap
state that appears above the superconducting transition in the underdoped materials is probably best regarded as
a doped version of a {\em paramagnetic} Mott insulator. Such an insulator has a `built-in' spin
(pseudo)gap.
Vexing questions however remain on the precise
theoretical connection between such a point of view and the actual occurence of antiferromagnetic order in the
undoped materials.

\begin{figure}
\includegraphics[width=4cm]{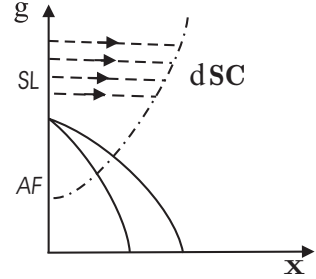}
 \caption{Zero temperature phase diagram showing the route from Mott insulating antiferromagnet to
d-wave superconductor (dashed-dot line). Horizontal axis refers to
doping and vertical axis refers to frustrating spin interactions that destabilize the Neel
state. The thesis of the spin liquid approach is that the intermediate and long scale physics of
the doped system may be fruitfully viewed as those of a doped spin liquid Mott insulator. Doping the
spin-liquid phase naturally leads to d-wave superconducting state.
(dashed lines).} \label{sf} \label{phase}\end{figure}

A clue to resolving this dilemma is provided by neutron scattering
experiments that reveal the existence of a sharp magnetic resonance
at $(\pi, \pi)$ in the doped superconductor
\cite{peakf,peaks,peakt}. An appealing interpretation of this
resonance is as a gapped version of the familiar magnon of a
proximate antiferromagnetic state\cite{csy,mrpns,vs}. Interestingly as the
doping is reduced, the resonance frequency goes down proportionately
to $T_c$\cite{edop,edopp}. This suggests that if the doped state is
to be viewed as a doped paramagnet, then the latter may at least be
connected to the Neel state by a second order transition (see Fig
\ref{phase}). Thus we are lead to search for quantum paramagnetic
states of spin-$1/2$ moments on a square lattice that are accessible
from the collinear Neel state by a second order transition.

It is important at this point to also review another old idea in
cuprate theory. Early work\cite{pwa,krs} suggested that the undoped
magnetically ordered Mott insulator is close to being disordered by
quantum fluctuations into a paramagnetic featureless ``spin liquid"
phase. Such a spin liquid state was postulated to have neutral
spin-$1/2$ spinon excitations and preserve all the symmetries of the
underlying microscopic spin Hamiltonian (including spin rotation).
Further it was argued that doping a spin liquid could possibly lead
to high temperature superconductivity\cite{pwa,krs}.

Despite its original appeal, this scenario was subsequently
questioned by a number of significant theoretical developments.
Calculations in a controlled large-$N$ expansion of quantum
Heisenberg spin models concluded that the natural result of
destruction of collinear Neel order was not a featureless spin
liquid but (for spin-$1/2$) a valence bond solid (VBS), which breaks
various lattice symmetries\cite{ReSaSuN}. (Here natural refers to
the possibility that the quantum paramagnet in question is
potentially separated from the Neel state by a second order
transition.) The VBS state also does not support fractionalized
spinon excitations. This was supported by a number of other indirect
arguments - for instance by studies of quantum dimer models on the
square lattice\cite{subirqdm}. It was shown however that destruction
of non-collinear Neel order could indeed lead to a fractionalized
spin liquid state that preserves all lattice
symmetries\cite{ReSaSpN}.

These calculations lead to the following folk wisdom (for a review see Ref. \onlinecite{subir}):
``{\em In two
spatial dimensions collinear ordered magnets naturally give way to
confined VBS paramagnets when disordered by quantum fluctuations
while non-collinear magnets naturally lead to spin
liquids}". As the magnetic ordering is undoubtedly
collinear in the cuprates this folk wisdom apparently spells doom
for the view of the cuprates as doped spin liquid paramagnetic Mott
states.

The purpose of the present paper is to revisit these issues. We will first argue that, if at all a
spin liquid based approach is to be pursued, experiments suggest
a certain kind of paramagnetic spin liquid state as natural candidate `parent' states of the doped cuprates. Next we
argue that the existing theoretical work does not rule out a direct second order transition between the Neel state
and this particular kind of spin liquid state. Finally we outline
in some detail a theory for just such a direct second order transition. Thus our work calls into question the folklore
described above and potentially frees
the spin liquid based approach to the cuprates from one of its theoretical criticisms.

Based on these results we will develop a qualitative picture of the
neutron resonance mode seen in experiments in the doped system and
its relationship with other aspects of the observed spin physics.
Our description will naturally unify two popular views of the
resonance mode - one as a soft mode associated with
antiferromagnetic long range order\cite{csy,mrpns,vs}, and the other as a
spin exciton formed as a triplet particle-hole collective mode of
fermionic quasiparticles\cite{liu,milmo,abchub,norman,bmanlee}.

We begin with experiments. It is by now quite clearly established
that the cuprate superconductors are $d$-wave paired and furthermore
have nodal BCS-like quasiparticles. The existence of the nodal
quasiparticles is theoretically significant. Indeed the possibility
of $d$-wave paired superconductors without nodal quasiparticles has
been much emphasized by Kivelson and coworkers (for a review see Ref. \onlinecite{kivrev}). It is therefore
of some interest to ask whether there exist paramagnetic Mott states
that already have gapless nodal excitations. Such a state then
builds-in enough of the spin physics seen in the experiments at
finite doping that it would be an attractive `parent' Mott insulator
as a basis for a theory of the underdoped cuprates.

Remarkably such states are known to
exist as stable quantum phases\cite{z2long,stable-u1} of quantum antiferromagnets
magnets on a two dimensional square lattice, at least within an appropriate large-$N$
expansion. In this paper we will
focus on one such state that has played a central role in some
previous theoretical
work\cite{affleck-marston-rapid,affleck-marston-prb,WLsu2,KL9930,RWspin,sentlee}
on the cuprate problem. This state - dubbed the $d$-wave RVB or
staggered flux (sF) spin liquid - is a quantum paramagnet that
nevertheless has gapless spin carrying excitations. Recent
theoretical work\cite{stable-u1} has established the stability of
such a state (in a suitable large-$N$ expansion). A low energy
description of the physics is usefully provided in terms of a theory
of gapless nodal fermionic spinons coupled minimally to a
fluctuating $U(1)$ gauge field. Despite this however there really is
no true quasiparticle description of the low energy
spectrum\cite{RWspin,su4}.

In this paper we are interested in exploring the quantum phase
transition between this algebraic spin liquid and the collinear Neel
state. Previous theoretical work providing the basis for the folk
wisdom mentioned above do not constrain the nature of this
transition. First the large-$N$ calculations of Ref.
\onlinecite{ReSaSuN} were based on a bosonic representation of the
spins. This representation is not well-suited to access the
algebraic spin liquid phase. Indeed it is tailor-made to access
either VBS phases or gapped spin liquids with bosonic spinons.
Arguments for the presence of VBS order in the paramagnet based on
quantum dimer models on the square latice\cite{subirqdm} also do not
help. Clearly the quantum dimer models are useful only for
paramagnets with a full spin gap - and hence will not be able to
access algebraic spin liquids. A more recent elegant
argument\cite{sachdev-park} attacks from the spin liquid side as
follows. First it supposes that to access Neel ordered states from
spin liquids, the latter must have bosonic spinons. Then the Neel
state is reached simply by condensing the bosons. Examining the
dispersion relation of a bosonic spinon in a spin liquid state
reveals that it generically has a minimum at an incommensurate
wavevector. Condensing such a spinon then naturally leads to
incommensurate spiral states and not to the simple collinear state.
A loophole in this argument is the supposition that it is
only a spin liquid state with bosonic spinons that can be proximate
({\em i.e.} separated by a second order transition) to a Neel state.
Indeed we will show in this paper that the algebraic spin liquid
state - which doesn't have bosonic spinons - can be connected to the
Neel state by a second order transition.

Our starting point is a mean field description of both the spin
liquid and antiferromagnetic phases. This will be done in terms of a
slave particle representation of the spin operator in terms of
neutral fermionic spinons. At this mean field level the spin liquid
state we study will have spinons that are paired into a $d$-wave
state. The resulting spinon dispersion has four Fermi points in the
Brillouin zone at which the spinons are gapless. Antiferromagnetism
is obtained in this representation as a spin density wave transition
of the spinons. This magnetic ordering produces a gap for the
spinons. Such a mean field description of the antiferromagnetic state was
first proposed by Hsu\cite{Hsu}. Similar ideas have subsequently been explored in a
number of publications - see Refs. \onlinecite{nl2,topth,bk,ho}.
However it is important to realize that the resulting mean field state is {not}
a conventional antiferromagnet. Though the spinons have acquired a
gap they still have not disappeared from the spectrum. Thus the mean
field description is apparently that of a {\em fractionalized}
antiferromagnet. For the particular mean field state studied in this paper this
problem is cured once fluctuations beyond the mean field are
included. Indeed we will argue that these fluctuations confine the
spinons in the magnetically ordered state.

\begin{figure}
\includegraphics[width=4.8cm]{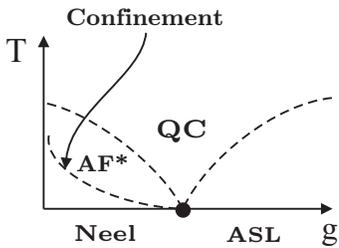}
\caption{Schematic phase diagram of quantum phase transition between
Neel and dRVB algebraic spin liquid phases.} \label{qc}
\end{figure}

Does the slave particle mean field description of the antiferromagnet still
have any physical meaning if the spinons are confined? An answer to this question is
provided by considering the magnetic state close to the transition to the spin liquid.
The critical point we describe for this transition has
some conceptual similarity with
the deconfined quantum critical points\cite{dqcp-science,dqcp-prb} studied recently. Specifically spinon variables
will already
prove useful in describing the critical point itself though they do not correspond to physical degrees of
freedom deep in the Neel phase. Furthermore there are two diverging length/time scales - one of which diverges as
a power of the other - as the transition is
approached from the Neel side. The physics is that of the conventional Neel state only at the very longest
scales. At length/time scales intermediate between the two diverging ones, the physics is correctly thought of
in terms of the mean field spin density wave state formed out of fermionic Dirac-like spinons. This state (though ultimately unstable)
describes an increasingly wider regime of intermediate scales close to the transition. Apart from the
spin waves expected from broken spin symmetry it also contains gapped spin-$1/2$ spinons coupled minimally to
a gapless spin-$0$ gauge boson (a ``photon"). Thus in this large intermediate scale regime the spinon mean field description of
the antiferromagnet is indeed the correct physical starting point.
Note that accessing the true critical theory requires keeping
fluctuations other than those of the Neel order parameter. This should hardly be a surprise given the
non-trivial nature of the paramagnetic phase.

The rest of the paper is organized as follows. In Section
\ref{meanf} we quickly outline the mean field theory of the spin
liquid and the transition to the Neel state. Next in Section \ref{bmft} we
go beyond mean field, by including gauge fluctuation for spinons and
dynamics for Neel field. We first briefly review the properties of the algebraic spin
liquid phase with particular emphasis on
precursor fluctuations of the magnetic ordering near the transition (Section \ref{sp}).
Then in subsection \ref{ms} we discuss the magnetic state and
show that it gets smoothly connected to the conventional Neel state
once monopole fluctuations are included. We very briefly discuss the connection to
projected wavefunction descriptions of the magnetism in undoped spin models in subsection \ref{proj}.
Next in Section \ref{ptgen}
we consider the phase transition. We show that within certain assumptions
the transition has conceptually similar structure to the deconfined critical points
of Ref. \onlinecite{dqcp-science}. We then present results for both the ASL phase and
the transition to the Neel state within an appropriate $\epsilon$ expansion in Section \ref{eps}.
In Section \ref{hitcimp}  we discuss implications of our results
for theories of cuprates before concluding in Section \ref{concl}. Various appendices contain technical details
of the calculations.

\section{Mean field theory}\label{meanf}
Consider a generic $SU(2)$ symmetric spin-$1/2$ model on a square lattice with predominantly antiferromagnetic short ranged interactions:
\begin{equation}
{\cal H} = J \sum_{<rr'>} \vec{S}_{r} \cdot \vec{S}_{r'} + \cdots
\label{eqn:spin-model}
\end{equation}
Here $J > 0$ (antiferromagnetic exchange), and the ellipsis
represent frustrating interactions that can be used to tune quantum
phase transitions. We will require that the full Hamiltonian be
invariant under $SU(2)$ spin rotations, time-reversal, and the full
space group of the square lattice. It is well known that the nearest
neighbour model has a Neel ordered ground state. Various
paramagnetic ground states can be accessed (in principle) by
appropriate frustrating interactions. As explained in the
introduction, here we will focus on a particular paramagnetic state
that is known as the dRVB algebraic spin liquid (also often referred
to as the staggered flux spin liquid). A mean field theory for this
state has been described several times in the literature and is
well-known\cite{affleck-marston-rapid}. First the spin is formally
rewritten as a bilinear of fermionic ``spinon'' operators
\begin{equation}
\vec{S}_{r} = \frac{1}{2} f^\dagger_{r\alpha} \vec \sigma_{\alpha
\beta} f^{\vphantom\dagger}_{r \beta} .
\end{equation}
Here $\alpha = 1,2$, corresponding to spin up/spin down fermions.  This is an exact rewriting when combined with the local constraint $f^\dagger_{\alpha} f^{\vphantom\dagger}_{\alpha} = 1$.  In the mean field approximation the
exact Hamiltonian is replaced by one quadratic in the spinon operators but with self-consistently determined parameters.
For the dRVB state, the mean field Hamiltonian takes the form
\begin{equation}
H_{sF} = - \sum_{<rr'>} \left((\chi_{rr'} + i\Delta_{rr'})f^{\dagger}_rf_{r'} + h.c\right)
\end{equation}
Here we take $r$ to belong to one sublattice of the square lattice.
So that $r'$ belongs to the opposite sublattice. The constants
$\Delta_{rr'} = +\Delta$ on horizontal bonds, and $-\Delta$ on
vertical bonds while $\chi_{rr'} = t$ on all bonds. This describes
fermionic spin-$1/2$ spinons on the square lattice with complex
hopping amplitudes such that there is a non-zero flux that is
staggered from plaquette to plaquette. Despite appearances, this
saddle point possesses the full symmetry of the microscopic model
including all lattice symmetries. (The apparent breaking of
translational symmetry is a gauge artifact). Recent work has
clarified the nature of fluctuations about this mean field state.
But in this present section, we will stick to the mean field
description and see how a transition to a Neel ordered state may be
described.

To access a Neel state we modify the dRVB mean field Hamiltonian by
adding a nearest neighbour antiferromagnetic interaction between the
spinons. Such an interaction will anyway be induced once
fluctuations beyond the mean field theory are considered. By
including it explicitly, we can induce a spin density wave ordering
of the fermionic spinons. We therefore consider
\begin{equation}
H = -\sum_{\langle rr' \rangle} (T_{rr'}
f_r^{\dag}f_{r'}+T^{*}_{rr'}
f_{r'}^{\dag}f_r)+\frac{1}{g}\sum_{\langle rr' \rangle}\vec S_r .
\vec S_{r'}
\end{equation}
Here $T_{rr'} = T = t + i\Delta$ on bonds as shown in Fig. \ref{sf}
and equals $T^* = t-i\Delta$ on other bonds as also shown in Fig
\ref{sf}. We now treat the $\frac{1}{g}$ term in a mean field
approximation. We look for a solution where $<\vec S_r> = \epsilon_r
N \hat{z}$ is non-zero (In mean field theory, $\vec N=\langle
\epsilon_r \vec S_r \rangle$ is constant. So we can choose it's
direction as $z$ direction). Here $\epsilon_r = (-1)^{x+y}$ is $+1$
on the A sublattice and $-1$ on the B sublattice. The mean field
Hamiltonian reads
\begin{equation}\label{hamiltonian}
H_{MF} = -\sum_{\langle rr' \rangle} (T f_r^{\dag}f_{r'}+T^{*}
f_{r'}^{\dag}f_r)-\frac{4N}{g} \sum_r\epsilon_r f_r^{\dag}\frac{
\sigma^z}{2}f_r
\end{equation}
The value of $N$ is to be determined self-consistently.
 We can diagonalize this Hamiltonian using a two site
unit cell as plotted in figure \ref{sf}. This gives the following
equation for energy eigenvalues and eigenstates:
\begin{equation}\label{evalue}
\left[\begin{array}{ccc} -\frac{4}{g} N\frac{\sigma^z}{2} & \varepsilon(\textbf{k})-i\Delta(\textbf{k})\\
\varepsilon(\textbf{k})+i\Delta(\textbf{k}) &
\frac{4}{g}N\frac{\sigma^z}{2}
\end{array}\right]\left[\begin{array}{ccc} f_1 \\ f_2
\end{array}\right]=E_\textbf{k}\left[\begin{array}{ccc} f_1 \\ f_2
\end{array}\right]
\end{equation}
\begin{figure}
\includegraphics[width=4.5cm]{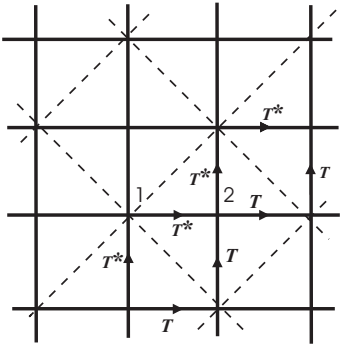}
\vspace{0.1in} \caption{Two-site unit cells (indicated by dashed
lines) used to diagonalize mean field Hamiltonian. } \label{sf}
\end{figure}
Heres, $f_1$ belongs to one sublattice and $f_2$ to the other one.
We have decomposed $T$ as $t-i\Delta$. $\varepsilon(\textbf{k})$ and
$\Delta(\textbf{k})$ are then defined as:
\begin{equation}\nonumber\begin{split}
\varepsilon(\textbf{k})&=-2t(\cos(k_y)+\cos(k_x))\\
\Delta(\textbf{k})&=2\Delta(\cos(k_x)-\cos(k_y))
\end{split}
\end{equation}
This gives spectrum of the two bands:
\begin{equation}\label{spectrum}
E^{\pm}_{\textbf{k}}=\pm\sqrt{(\frac{2N}{g})^2+\varepsilon(\textbf{k})^2+\Delta(\textbf{k})^2}
\end{equation}
Now with this in hand we can derive the self-consistency equation
for $N$:
\begin{equation}
\langle -\frac{\partial H_{MF}}{\partial (\frac{4N}{g})}
\rangle=\sum_r\langle\epsilon_r f^{\dag}_r \frac{
\sigma^z}{2}f_r\rangle =L^2 N
\end{equation}
Where $L$ is the linear system size. Using the spectral function of
the lower band (we consider $T=0$) this gives:
\begin{equation}\label{sec}
g\ N=\frac{N}{L^2}\sum_{\textbf{k}}\frac{1} {\sqrt{(\frac{2
N}{g})^2+\varepsilon(\textbf{k})^2+\Delta(\textbf{k})^2}}
\end{equation}
This equation has a trivial solution $N=0$, but it is obvious from
(\ref{spectrum}) that non-zero solution, if it exists, has lower
energy. So the system has two phases, achieved by tuning the value
of $g$. The critical value, $g_c$ is given by:
\begin{equation}
g_c=\frac{1}{L^2}\sum_{\textbf{k}}\frac{1}
{\sqrt{\varepsilon(\textbf{k})^2+\Delta(\textbf{k})^2}}
\end{equation}
for $g>g_c$ there is no non-zero solution and so $N=0$. But for
$g<g_c$ we have $N\neq 0$. From (\ref{sec}) we can also derive the
behavior of $N$ at critical point within mean-field:
\begin{equation}
N\propto \left\{ \begin{array}{ccc} 0 &\ for\ \  g>g_c \\
g_c-g &\ for\ \  g<g_c
\end{array}\right.
\end{equation}
Note that in the magnetic phase the non-zero $N$ induces a gap to
the spinons.

To study low energy properties, we will set up a continuum effective
theory. In the next section we will include fluctuations in this
continuum field theory. In the spin liquid state, the spectrum
consists of two Fermi points, located at $k_x=k_y=\frac{\pi}{2}$ and
$k_x=-k_y=-\frac{\pi}{2}$ (at these points
$\varepsilon(\textbf{k})=\Delta(\textbf{k})=0$), in the reduced
Brillouin zone. There are gapless spinon excitations near these
nodes with a Dirac-like linear dispersion. A low energy description
of the spin liquid is then possible in terms of a continuum field
theory of massless Dirac spinons (a brief review that helps fix
notation is in appendix \ref{Dirac}). To study the magnetic
transition described above within this continuum field theory, we
need to introduce a `mean field' that couples to the $(\pi,\pi)$
component of the physical spin density. The resulting action takes
the form
\begin{equation}\label{maction}
S_m=\int d^2xd\tau \bar{\psi}(-i\gamma^{\mu}\partial_{\mu}+i\lambda
\ \mu^z N.\frac{\sigma^z}{2})\psi
\end{equation}
In this representation, $\psi$ consists of four two-component Dirac
fields. The four Dirac fields arise from the presence of two
physical spin species together with two pairs of nodes. The Pauli
matrices $\vec \sigma$ act on the spin index while the $\vec \mu$
are Pauli matrices acting on the node index. It is readily seem that
the combination $\frac{i}{2}\bar{\psi}\mu^z \vec \sigma \psi$ is
precisely the continuum form of the physical spin density near
$(\pi, \pi)$. In the mean field theory $N$ is to be determined
self-consistently. The coupling  $\lambda$ is proportional to
coupling $\frac{1}{g}$ and from now on the momenta are considered
with respect to the nodes. As expected a non-zero $N$ gaps out the
Dirac spinons. This gap vanishes upon approaching the phase
transition to the spin liquid. The inverse of this gap determines a
diverging length scale - within the mean field theory this length
scale describes the decay of the {\em connected} part of the spin
correlations near $(\pi, \pi)$. This may be seen by a direct
calculation (described in appendix \ref{feynman}) which gives:
\begin{equation}\label{corm}
e^{i\vec Q.\vec r} \langle S_i(0) S_j(r) \rangle_c \propto
\frac{e^{-\frac{r}{\xi}}}{r^4}(1+\frac{r}{\xi})\ \delta_{ij}
\end{equation}
So the connected correlation for $r\ll\xi$ is a power-law decaying function
with fourth power of $r$. For $r\gg\xi$ it is a exponential decaying
function, with correlation length $\xi$ and a pre-factor which
decays as the third power of $r$. The correlation length at critical
point diverges as:
\begin{equation}\label{corlm}
\xi\propto\frac{1}{|\lambda-\lambda_c|} \sim \frac{1}{|g - g_c|}
\end{equation}

\section{Beyond mean field theory}
\label{bmft}
In this Section we consider the effects of fluctuations
beyond the mean field theory described above. In the spin liquid
phase far from the magnetic transition the crucial fluctuations are
those associated with the phase of the spinon hopping parameter.
These are to be thought of as gauge fluctuations associated with a
(compact) $U(1)$ gauge field that is coupled minimally to the
spinons. Recent work has shown that the dRVB spin liquid is stable
to such gauge fluctuations\cite{stable-u1} (at least within a
systematic large-$N$ expansion where $N$ is the number of Dirac
spinons). Through out this paper we will assume that this stability
persists to the physically relevant case $N = 4$. The low energy
theory of the resulting phase is described by massless $QED$ in
three space-time dimensions:
\begin{equation}
S=\int d^2x d\tau\ \{ -\bar{\psi}[i\gamma^{\mu}(\partial_{\mu}  +i\
e\ a_\mu)]\psi +
(\epsilon_{\mu\nu\kappa}\partial_{\nu}a_{\kappa})^2\}
\end{equation}
Here $a_{\mu}$ is a fluctuating $U(1)$ gauge field which may be taken to be non-compact
at the low energies.
This theory flows to a conformally invariant
fixed point. Various physical quantities have non-trivial power law correlations at the resultant
spin liquid fixed point\cite{wasl}. In particular the $(\pi, \pi)$ spin correlator decays as a power law:
\begin{equation}
e^{i\vec Q.\vec r} \langle \vec S(0) . \vec S(r) \sim \frac{1}{r^{2\Delta}}
\end{equation}
The exponent $\Delta$ is not known - a rough estimate from projected
wavefunctions\cite{proj1,proj2} gives $\Delta \approx 0.75$. The
dynamical spin correlations at $(\pi, \pi)$ in the scaling limit
follow straighforwardly from the relativistic invariance of the
field theory above. For the full zero temperature dynamical spin
susceptibility, we have
\begin{equation}
\chi_{SL}''(q, \omega) \sim \frac{1}{\omega^{2-\eta}}F\left(\frac{\omega}{vq}\right)
\end{equation}
Here $\vec q$ is the deviation of the wavevector from $\vec Q =
(\pi, \pi)$, $F$ is a universal scaling function, and $v$ is a
non-universal spinon velocity associated with the nodal Dirac
dispersion. The exponent $\eta$ is the anamolous dimension of the
staggered spin and is related to $\Delta$ through $2\Delta =
1+\eta$. Due to the power law spin correlations, this spin liquid
phase has been dubbed as ``algebraic spin liquid" (ASL)\cite{wasl}.
Note that the spinons are not good quasiparticles at low energies in
the spin liquid phase. Indeed there presumably is no quasiparticle
description of the spectrum (rather like at interacting quantum
critical points). Nevertheless the field theory above in terms of
spinons provides a useful description of the system.

A remarkable feature of the dRVB algebraic spin liquid phase is the
emergence of a huge global symmetry group characterizing the low
energy fixed point. The low energy theory has an $SU(4)$ symmetry
corresponding to free unitary rotations between the four Dirac
species. In addition the irrelevance of space-time monopoles at low
energies implies a non-trivial global $U(1)$ symmetry associated
physically with the conservation of internal magnetic flux. Ref.
\onlinecite{su4} studied a number of consequences of the $SU(4)$
symmetry. In particular it showed that several other competing order
parameters had the same power law correlators as the Neel vector -
these include the order parameter associated with the
columnar/plaquette VBS orders.

Near the transition to the antiferromagnetic state, we must treat
the Neel field introduced in the previous section as a fluctuating
vector $\vec N$. Further upon integrating out high energy spinons
({\em i.e.} ones far away from the nodes), this $\vec N$ field will
develop some dynamics of its own.  The resulting action takes the
form
\begin{equation}\label{action}\begin{split}
S=\int d^2x d\tau\ &\{  -\bar{\psi}[i\gamma^{\mu}(\partial_{\mu}
+ie\ a_\mu)]\psi+i\lambda \bar\psi(\mu^z\  \vec N.\frac{\vec
\sigma}{2})\psi+\\
&(\epsilon_{\mu\nu\kappa}\partial_{\nu}a_{\kappa})^2
 +\frac{(\partial_{\mu}\vec N)^2}{2}+r\frac{(\vec
N)^2}{2}+\frac{u}{4!}((\vec N)^2)^2\}
\end{split}
\end{equation}

In writing this action we have ignored anisotropies in the spinon
velocities at the Dirac node and any difference between the
velocities of spinon and $N$ fields. Later we will show that all of
these velocity anisotropies are irrelevant (if small) at the
critical fixed point between the dRVB ASL and Neel states (see
\ref{velocity}).

In the presence of the coupling to the $\vec N$ field, the action no
longer has full $SU(4)$ global symmetry. An $SU(2)$ subgroup -
corresponding to physical spin rotations - is still obviously a
symmetry. This involves an $SU(2)$ spin rotation of the $\psi$ field
together with an $O(3)$ rotation of the $\vec N$ vector. In
addition, the global transformation
\begin{equation}
\psi \rightarrow e^{i\theta \mu^z} \psi
\end{equation}
with $\theta$ a constant is also a symmetry. This is a $U(1)$
subgroup of the full $SU(4)$ symmetry. Thus the action has a global
$SU(2) \times U(1)$ symmetry apart from the $U_{flux}(1)$ associated
with the gauge flux conservation. The extra $U(1)$ symmetry that
survives from the full $SU(4)$ has the consequence that fermion
bilinears such as $\bar{\psi}\mu_x\psi, \bar{\psi}\mu_y\psi$ can be
freely rotated into another. In the original spin model, these
operators transform identically to the VBS order parameter. The
$U(1)$ symmetry then implies that the columnar and plaquette order
parameters can be rotated into one another, and hence will have
identical correlations.

Let us now study some general aspects of the two phases, near the
phase transition.

\subsection{Precursor fluctuations in the spin liquid}\label{sp}
We will first consider the precursor fluctuations of the magnetic ordering in the spin liquid side.

\begin{figure}[htp]
\includegraphics[height=2.9cm]{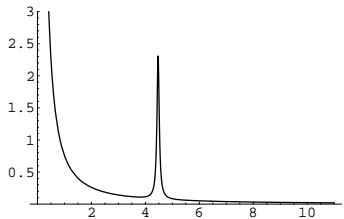}
\caption{Dynamical spin susceptibility at ($\pi$,$\pi$)  in the
spin liquid phase in $\lambda$=0 limit} \label{suscep}\end{figure}

First consider the limit $\lambda = 0$. Then the $\vec N$ field
decouples from the spinon-gauge sector.  It is instructive to think
about the spectral function for the $(\pi, \pi)$ spin correlations
in the spin liquid phase in this limit. It is simply a sum of two
pieces as shown in Fig. \ref{suscep} - a diverging power law coming
from the ASL, and a sharp delta function peak coming from the $N$
fluctuations. Now consider turning on a small non-zero $\lambda$.
The low frequency divergence of the spin susceptibility will be
unaffected by the coupling to the fluctuating $\vec N$ field (this
follows from the assumed stability of the ASL fixed point). However
the delta function peak coming from the $\vec N$ field will now be
broadened due to decay into two spinons. The broadening may be
described within a simple RPA approximation (for details see
appendix \ref{rpa}).

\begin{figure}[htp] \centering
\begin{center}
\vspace{0.1in}\subfigure[]{\label{sca}
\includegraphics[height=2.9cm]{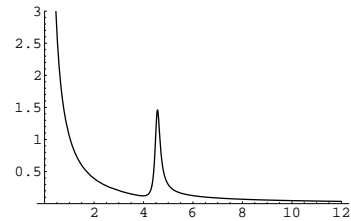} \vspace{0.1in}
} \subfigure[]{
\includegraphics[height=2.9cm]{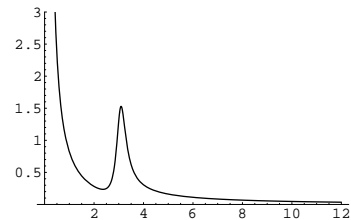}
\vspace{0.1in} \label{scb}} \subfigure[]{
\includegraphics[height=2.9cm]{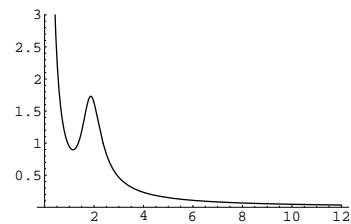}
\vspace{0.1in} \label{scc}}
 \caption{Dynamical spin susceptibility at $(\pi,\pi)$ after turning on a nonzero value for $\lambda$.
 The plot shows the change upon approaching
 the transition from \ref{sca} to \ref{scc}. Note the softening of the $\vec{N}$ peak, as the transition occurs.}
\label{sc}\end{center}
\end{figure}

In Fig \ref{sc} we plot the dynamical spin susceptibility at $(\pi,
\pi)$ as the phase transition is approached (by decreasing $r$).
Note that as expected, the peak coming from the $\vec N$
fluctuations ``softens" on approaching the transition.

\subsection{Magnetic state}
\label{ms} When the parameter $r$ is sufficiently negative the $\vec
N$ field will condense leading to magnetic long range order. In the
continuum field theory this may be viewed as a spin density wave
state that arises out of the dRVB ASL. In this subsection we will
argue that contrary to naive expectations, it is in fact a different
state from a conventional Neel state - rather it is a fractionalized
antiferromagnet in the same spirit as that studied in Refs.
\onlinecite{nl2,z2long}. This apparent problem will be cured once we
include the effects of monopole fluctuations (ignored so far). We
will show that this fractionalized antiferromagnet evolves at long
length/time scales into the conventional Neel antiferromagnet.

Consider first the description of the Neel ordered state within the
mean field theory developed in Section \ref{meanf}.  The mean field
spectrum consists of {\em gapped} spin-$1/2$ spinons\footnote{Note
that if $\vec N = N_0 \hat{z}$, then $S^z$ is a good quantum number
and can be used to label the states. The spinons have $S^z = \pm
1/2$.}. It is important to realize that the spinons are merely
gapped - they however have not disappeared from the spectrum. Now
consider including fluctuations as in Section \ref{bmft}. The
important fluctuations are those associated with slow rotation of
the direction of the Neel order parameter $\vec N$ (spin waves) and
those asociated with the phase of the fermion hopping $T$ (gauge
fluctuations). These are both conveniently discussed within the
continuum theory in Eqn. \ref{action}, that obtains close to the
critical point. Integrating out the gapped spinons, we may obtain an
effective action for the spin waves and the gauge fluctuations. To
quadratic order in both the transverse component of the Neel vector
$\vec N_{\perp}$ and the gauge field $a$, we get
\begin{equation}
\label{eftord} S_{eff} = \int d^3x
\frac{\rho_s}{2}\left(\partial_{\mu} \vec N_{\perp} \right)^2 +
\frac{g}{2}\left(\epsilon_{\mu \nu \kappa} \partial_{\nu} a_{\kappa}
\right)^2
 + .......
\end{equation}
where the ellipses refer to higher order terms that are unimportant
at low energies. The first term describes the expected gapless spin
wave excitations. The second term describes a gapless linear
dispersing ``photon" associated with the gauge fluctuations. This
extra gapless mode provides a sharp low energy distinction between
this Neel state (as described so far) and the conventional one. The
gapless photon mode is minimally coupled to the gapped spinons - the
presence of gapped spinons serves as another distinction with the
conventional Neel state. Thus the antiferromagnet state is to be
characterized as a ``fractionalized antiferromagnet" with a $U(1)$
gauge structure. Following the notation of Ref. \onlinecite{nl2}, we will
dub it $U(1)$ $AF^*$.

Let us now include monopole fluctuations. In this magneticaly
ordered phase, the low energy gauge action is that of free Maxwell
theory in $2+1$ dimensions. Then standard arguments show that the
monopoles are strongly {\em relevant}. Thus the $U(1)$ $AF^*$ state
(in two dimensions) is ultimately unstable to monopole
proliferation. The result is to gap out the photon mode and cause
confinement of all objects that carry non-zero gauge charge. In
particular it implies that the spinons (which survived as gapped
excitations when monopoles were ignored) will now be confined and
disappear from the spectrum. The resulting state is thus simply
smoothly connected to the conventional Neel state. Thus including
monopole fluctuations cause an instability of the unconventional
$U(1)$ $AF^*$ state toward the conventional Neel state.

\subsection{Projected wavefunctions}
\label{proj} Before continuing we digress briefly to make contact
with the large body of work on Gutzwiller projected superconducting
wavefunctions (see Ref. \onlinecite{proj2} and references therein),
as a route to implementing RVB ideas. Of interest to us, will be
studies on projected $d$-wave BCS states and their variants. In the
slave particle description, a useful guess for a prototypical
wavefunction for a state is obtained by taking the mean field state
and projecting it onto the space of physical states. At half-filling
this is equivalent to doing a Gutzwiller projection on the mean
field state. According to this prescription, a guess for the
wavefunction of the dRVB algebraic spin liquid will simply be
\begin{equation}
|dRVB> = P_G |dBCS>
\end{equation}
where $|dBCS>$ is the mean field ground state of a $d$-wave
superconductor at half-filling with just nearest neighbour hopping
and pairing on the square lattice. Correspondingly, a guess for the
wavefunction of the magnetic state, as we have obtained it, would
simply be
\begin{equation}
|AF> = P_G |dBCS + SDW>
\end{equation}
The preprojected state on the right simply has spin density wave
order at $(\pi,\pi)$ coexisting with the $d$-wave superconductivity.
Such wavefunctions have been studied numerically\cite{gros} and are
known to have excellent energy for the nearest neighbour Heisenberg
model. From our considerations in the previous section, we would
expect that this wavefunction is a prototype for a {\em confined}
antiferromagnet with no finite energy spinons. Some support for this
expectation comes from the work of Ref. \onlinecite{leeshih} which
studied the properties of a single hole in that state. The
quasiparticle residue was found to be non-zero consistent with that
expected in a confined antiferromagnet.

\section{Phase transition: Generalities}
\label{ptgen}

Let us now consider the phase transition between the dRVB ASL and the Neel state. In the limit $\lambda = 0$,
the $\vec N$ vector fluctuations are decoupled from the ASL and the magnetic transition is simply
in the universality class of the usual $O(3)$ fixed point in $D = 3$ space-time dimensions.
Note that
in this limit mean field theory predicts that the Neel order parameter vanishes with exponent $\beta = 1/2$
on approaching the transition.
What is the effect of turning on a weak $\lambda$ at this decoupled transition? First note
that in the mean field theory of Section \ref{meanf} we found that the Neel order parameter vanished with
exponent $\beta = 1$ clearly different from the $\lambda = 0$ limit. Thus a
non-zero $\lambda$ already changes the answers within mean field theory. More generally the
effects of a weak $\lambda$ may be assessed by
considering the renormalization group flow of $\lambda$ at the decoupled fixed point. We have
\begin{equation}
\frac{d\lambda}{dl} = (D - \Delta_N - \Delta)\lambda
\end{equation}
where $D = 3$ is the space-time dimension, $\Delta_N$ is the scaling
dimension of the $N$ field at the $D = 3$ $O(3)$ fixed point, and
$\Delta$ is the scaling dimension of the spin operator near $(\pi,
\pi)$ at the dRVB ASL fixed point. Here $l$ is the usual logarithmic
renormalization scale. We have $\Delta_N = \frac{1+ \eta_N}{2}
\approx \frac{1}{2}$, and $\Delta = \frac{1+\eta}{2}$. Thus $D -
\Delta_N - \Delta \approx 2 - \frac{\eta}{2}$. With the rough
estimate $\eta \approx 0.5$, we find that $\lambda$ is strongly
relevant at the decoupled fixed point. Thus the true critical
behavior will involve strong coupling between the $\vec N$ field and
the spinons of the ASL. In Section \ref{eps}, we will study this
critical behavior in a controlled $3-\epsilon$ dimension. A very
similar field theory where a fluctuating $O(3)$ vector field was
coupled to massless Dirac fermions was studied many years ago by
Balents et al\cite{ndllquid} in a different physical context. The
main difference between the theory of Balents et al and the action
in Eqn. \ref{action} is the presence of the gapless gauge fields in
the latter. We will see that this modifies the universality class of
the transition from that in Ref. \onlinecite{ndllquid}.

\begin{figure}[htp]
\includegraphics[width=5cm]{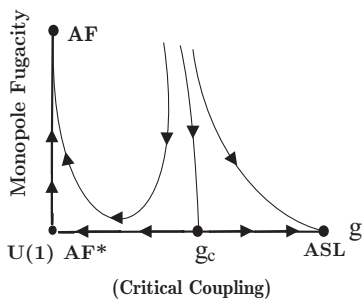}
\caption{Renormalization flow diagram near the critical fixed point. The vertical axis is the
 monopole fugacity; the horizontal axis is a coupling $g$ which describes the strength of the
 short range part of the spinon interaction.}
\label{fld}\end{figure}

What about monopole fluctuations at this critical point?  Let us
first review the situation in the paramagnetic algebraic spin liquid
state. Recent work has argued\cite{stable-u1} that when the number
of Dirac species $N$ is sufficiently large ({\em i.e.} bigger than
some $N_c$) the monopoles are formally irrelevant at the ASL fixed
point\footnote{It is at present not known what the value of $N_c$ is
though existing numerical work suggests the bound $N_c < 8$. For the
$SU(2)$ magnet studied in this paper we have $N = 4$ and have simply
assumed that $N_c < 4$.}.  In the large-$N$ expansion, the monopole
scaling dimension will be $o(N)$ both at the ASL fixed point and at
the critical fixed point - thus at least for large enough $N$ the
monopoles are irrelevant at the critical fixed point as well. In
this paper we will make the crucial assumption that this irrelevance
continues to hold at $N = 4$, {\em i.e.} for  $SU(2)$ spin models
(see figure \ref{fld}).

\begin{figure}[htp]
\includegraphics[height=1.3cm]{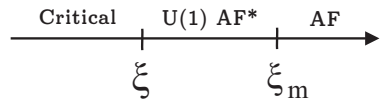}
\caption{Crossover length scales in the magnetic state close to the transition to the spin liquid.
The shorter length scale $\xi$ describes the crossover from the critical state to the fractionalized antiferromagnet.
The longer scale $\xi_m$ is where this exotic antiferromagnet crosses over to the conventional Neel state
through confinement. Both scales diverge near the critical point but $\xi_m$ diverges faster than $\xi$.}
\label{lss}\end{figure}

With this assumption the monopole fugacity is irrelevant at the
critical fixed point (and the paramagnetic ASL fixed point) but
relevant at the ordered fixed point of the continuum field theory in
Eqn. \ref{action}. In renormalization group language, the monopole
fugacity is a dangerously irrelevant coupling. The length scale
$\xi_m$ at which the photon gets gapped (which may loosely be dubbed
the ``confinement scale'') in the magnetic side may be estimated as
follows. Let the monopole scaling dimension at the critical point be
$\Delta_m > 3$. Upon scaling out of the critical region in the
ordered state to the correlation length scale $\xi$, the monopole
fugacity will renormalize to $z_{\xi} \sim \xi^{3-\Delta_m}$. It is
beyond this scale $\xi$ that the action in Eqn. \ref{eftord} starts
applying. In the free Maxwell theory that obtains beyond $\xi$, the
monopole fugacity grows. A standard matching argument now gives
$\xi_m \sim \xi^{(\Delta_m -1)/2}$. Thus $\xi_m$ diverges faster
than $\xi$. The physics on scales $\xi \ll L \ll \xi_m$ is that of
the fractionalized antiferromagnet $U(1)$ $AF^*$. It is only at the
longest scales $L \gg \xi_m$ that the conventional Neel behavior is
obtained (fig. \ref{lss}).

\section{\protect{$\epsilon$} expansion for critical
properties}\label{eps} In this section we will show how the
structure of the critical fixed point can be studied in a formal
expansion near three space dimensions. Some care is necessary in
dealing with the Dirac matrices in arbitrary dimension. But here
following Ref. \onlinecite{ndllquid}, we will sidestep this issue.
We will perform calculations in a perturbative expansion of the
coupling constants in $ d = 2$, take the traces over the Dirac
matrices and then finally, in evaluating momentum integrals, set $d
= 3-\epsilon$. As a warm-up we first describe the ASL fixed point
within this $\epsilon$ expansion.
\subsection{ASL in the \protect{$\epsilon$} expansion}
Before starting the \protect{$\epsilon$} expansion studies for the
full theory, we study ASL phase with this approach. To do that we
consider action (\ref{action}) and assume $N=0$. Then it reduces to
the usual $QED_3$ action. The flow equation for this theory is well
known\cite{justin} (and also calculated in appendix \ref{feynman}):
\begin{equation}\label{qed}
\beta_{e^2}  =  \epsilon e^2-\frac{16}{3}\frac{e^4}{(4\pi)^2}
\end{equation}
Here $\epsilon = 3-d$.
This flow equation indicates the presence of nontrivial fixed point
of order $\epsilon$ at:
\begin{equation}
e^{*2} =  {3\pi^2} \epsilon
\end{equation}
 This is the pure ASL fixed point. The microscopic derivation of the continuum field theory for the
spin liquid allows for a velocity anisotropy between the two spinon
nodes. This anisotropy was found to be irrelevant in the large-$N$
limit of $QED_3$\cite{aniso1,aniso2,su4}. Here we examine its fate
within the $\epsilon$ expansion. Direct calculation (appendix
\ref{velocity}) shows that
$\beta_{\delta}=-\delta\frac{14}{3(4\pi)^2}\ e^{*2}=-\frac{7}{8}\
\delta\ \epsilon$, where $\delta$ measures the velocity anisotropy
({\em i.e.} $\delta=0$ corresponds to isotropic $QED_3$). So that it
is irrelevant at $o(\epsilon)$. Combined with the large-$N$ result,
this is strong evidence for its irrelevance at the physically
relevant ASL fixed point for $N = 4$ in two space dimensions. As
noted in previous papers\cite{su4} this irrelevance implies that the
$N =4$ ASL fixed point has global $SU(4)$ symmetry corresponding to
free unitary rotation between the four Dirac species.

Finally we can examine the scaling of gauge neutral fermion
bilinears (such as the Neel vector) in the $\epsilon$-expansion.
This is conveniently done by adding a source term that couples to
such a bilinear, and calculating the one loop correction to the
corresponding vertex (see appendix \ref{feynman}). We find that the
$(\pi, \pi)$ component of the spin has scaling dimension $\Delta =
3-1.94 \epsilon$. Setting $\epsilon = 1$ gives the estimate $\Delta
\approx 1.06$. This implies extremely slow decay of the
corresponding correlator. This estimate may be compared with that
from the $1/N$ expansion directly in $d = 2$ which gives $\Delta
\approx 1.54$. Thus both expansions give slow decay for the Neel
correlations that are strongly enhanced compared to the mean field
results.

\subsection{Critical fixed point}\label{cfp}
In this section we study the critical point within the $\epsilon$ expansion by
including the fluctuating $\vec N$ field.
>From the action (\ref{action}) we have three types of vertices and
eleven different one loop diagrams. At one loop level, as derived in
appendix \ref{feynman}, we get the following set of flow equations:
\begin{eqnarray}
\beta_{e^2} & = & \epsilon e^2-\frac{16}{3}\frac{e^4}{(4\pi)^2} \\
\beta_{\lambda^2} & = & \epsilon \lambda^2
-\frac{10}{(4\pi)^2}\lambda^4+\frac{10}{(4\pi)^2}e^2 \lambda^2 \\
\beta_u & = & \epsilon
u-\frac{11}{3}\frac{u^2}{(4\pi)^2}-16\frac{\lambda^2 u}{(4\pi)^2}+96\frac{\lambda^4}{(4\pi)^2} \\
\label{mass}\beta_r & =
&(2-\frac{5}{3}\frac{u}{(4\pi)^2}-8\frac{\lambda^2}{(4\pi)^2})r
\end{eqnarray}
Note that flow equation  for electric charge is the same as usual
quantum electrodynamics. In fact, gauge invariance dictates this
form\cite{ps,justin} (to keep the form $e\ a_{\mu}$ invariant under
RG flow we need to have $Z_e/\sqrt{Z_a}=1$). From these we can get
the following fixed points:
\begin{eqnarray}
e^{*2} & = & 0 \\ \lambda^{*2} & = & \frac{8\pi^2}{5}\epsilon \\
u^* & = & \frac{384\pi^2}{55}\epsilon
\end{eqnarray}
This fixed point describes the transition in the absence of the
gauge field and was first discussed by Balents et al\cite{ndllquid}.
Our calculations at this fixed point matches this previous work
which therefore provides a useful check. As another check at this
fixed point, if we consider our theory, with the three component
$\vec{N}$ field replaced by a scalar field $\phi$, it represent a
Yukawa like theory, which has been studied in Ref.
\onlinecite{justin}. Our results then can be partially checked
against these previous calculations. The full flow equations admit
another fixed point located at:
\begin{eqnarray}
e^{*2} & = & {3\pi^2} \epsilon\\
\lambda^{*2} & = & \frac{23\pi^2}{5} \epsilon\\
u^* & = & \frac{12 \pi^2}{55}(-36+\sqrt{12934})\epsilon
\end{eqnarray}
It is readily checked that the $e^2 = 0$ fixed point is unstable
towards this one. Thus the presence of the gauge field has changed
the universality class of the transition. Here we should also note
that the at the one loop level, calculation in appendix
\ref{velocity} shows that at this fixed point, the velocity
anisotropy is irrelevant at $o(\epsilon)$. So this fixed point is
also stable against small velocity anisotropy. Now using the flow
equation for $r$ (Eqn. \ref{mass}), it is easy to extract exponent
$\nu$ for this fixed point:
\begin{equation}
\frac{1}{\nu}  =  2-4.07\epsilon
\end{equation}
Note that simply setting $\epsilon = 1$ gives an unphysical answer.
This is a signal that the leading order $\epsilon$ expansion is not
quantitatively very accurate in estimating scaling dimensions in two
space dimensions. Despite this the $\epsilon$ expansion is useful to
describe the structure of the fixed points and the trends of the
various exponents.

It is very interesting to ask about the behavior of the staggered
spin correlations ({\em i.e.} near $(\pi, \pi)$) at this critical
point. Naively there are two different physical operators that have
the same symmetries as the staggered spin: the vector $\vec N$ and
the fermion bilinear $\vec N^z_A = \bar{\psi}\mu_z\vec \sigma \psi$.
Thus in writing down an expression for the staggered spin in terms
of the fields of the continuum theory, we must include both
contributions:
\begin{equation}
e^{i\vec Q. \vec x} \vec S(\vec x) \sim c_1 \vec N + c_2\vec N^z_A +...
\end{equation}
The ellipses refer to other operators with larger scaling dimension that also have the same symmetries
as the staggered spin. The coefficients $c_{1,2}$ are non-universal. Now consider the scaling of the
staggered spin. If $\lambda = 0$, then the fermion bilinear and $\vec N$ scale independently.
Near $d = 3$, and with the available estimate of the scaling dimension of the fermion bilinear at the ASL,
it is readily checked that $\vec N$ has the lower scaling dimension. Hence the long distance decay of the staggered
spin correlations will be determined by $\vec N$ in the $\lambda = 0$ limit. What happens when $\lambda$ is non-zero as
at the non-trivial fixed point above in the $\epsilon$ expansion? It is expected that the true scaling fields will be
some linear combinations of $\vec N$ and $\vec N^z_A$ which will have the form
\begin{eqnarray}
\vec \phi_1 & \sim & A_1\vec N + A_2\vec N^z_A \\
\vec \phi_2 & \sim & B_1\vec N + B_2\vec N^z_A
\end{eqnarray}
These fields will have scaling dimension $\Delta_{1,2}$ with (by definition) $\Delta_1 < \Delta_2$ so that the long distance decay will be dominated by $\vec \phi_1$.
The coefficients $A_{1,2}$ and $B_{1,2}$ will be determined by the fixed point theory.
At $o(\epsilon)$ we expect $A_1 \sim o(1), A_2 \sim o(\epsilon), B_1 \sim o(\epsilon), B_2 \sim o(1)$.
Thus to obtain the scaling dimensions $\Delta_{1,2}$ to $o(\epsilon)$ we can ignore the `mixing' terms
$A_2, B_1$, and simply calculate the anamolous dimension of $\vec N$ and $\vec N^z_A$.

The exponent $\eta$ for $N$ field (which determines the scaling dimension) is easily calculated from the
 field renormalization coefficient $Z_N$:
\begin{equation}
Z_N=1-\frac{8}{(4\pi)^2}\frac{\lambda^2}{\epsilon}
\end{equation}
$\eta$ is then given by coefficient of $\frac{1}{\epsilon}$ in
$Z_N$:
\begin{equation}
\eta=2.3 \epsilon
\end{equation}
so that $\Delta_1 = 1 + 0.65\epsilon$. The dimension $\Delta_2$ is
readily calculated as in our discussion of the ASL above. We find
$\Delta_2 = 3-1.65\epsilon$.

It is straightforward to determine the scaling dimension of all the
fermion bilinears related to $\vec N^z_A$ by $SU(4)$ rotations.
These are listed in Table \ref{bil}; the corresponding Feynman
diagrams are in Appendix \ref{feynman}. The absence of $SU(4)$
symmetry at the critical fixed point implies that these bilinears
mostly all have different scaling dimensions. Some weak constraints
follows from the $U(1)$ subgroup of the $SU(4)$ that remains
unbroken. For instance it implies that $N^x_A, N^y_A$ have the same
scaling dimension. As emphasized before physically this implies
identical scaling of the plaquette and columnar VBS order parameters
at this critical point.

\begin{table}[htp]
\begin{ruledtabular}
\begin{tabular}{c | c | c }
    $\text{  Field}$ & $\text{  Spin Model  }$ & $\text{Scaling}$ \\
$\text{theory}$ & \ & $\text{dimension}$
\\
    \hline
    $\vec{N}^x_A \text{ , } \vec{N}^y_A$ &
    $(-1)^{r_x + 1} \vec{S}_{\br} \times \vec{S}_{\br + \vec{y}}\text{ , }
    (-1)^{r_y} \vec{S}_{\br} \times \vec{S}_{\br + \vec{x}}$ & 3-0.5$\epsilon$\\
    \hline
    $\vec{N}^z_A$ & $(-1)^{r_x + r_y} \vec{S}_{\br}$ & 3-1.65$\epsilon$ \\
    \hline
    $\vec{N}_B$ & $(-1)^{r_x + r_y} \big[ (\vec{S}_1 + \vec{S}_3)(\vec{S}_2 \cdot \vec{S}_4)$ \\
        & $\qquad\qquad + (\vec{S}_2 + \vec{S}_4) (\vec{S}_1 \cdot \vec{S}_3) \big]$ & 3-1.65$\epsilon$ \\
    \hline
    $N^x_C \text{ , } N^y_C$ & $(-1)^{r_y} \vec{S}_{\br} \cdot \vec{S}_{\br + \by} \text{ , }
                         (-1)^{r_x} \vec{S}_{\br} \cdot \vec{S}_{\br + \bx}$ & 3-2.8$\epsilon$ \\
    \hline
    $N^z_C$ & $\big[ \vec{S}_1 \cdot (\vec{S}_2 \times \vec{S}_4) - \vec{S}_2 \cdot (\vec{S}_3 \times \vec{S}_1)$ \\
    & $ \,\,\,\,\,\,\,\,\,\,\, + \, \vec{S}_3 \cdot (\vec{S}_4 \times \vec{S}_2) - \vec{S}_4 \cdot (\vec{S}_1 \times \vec{S}_3) \big]$ & 3+0.65$\epsilon$\\
    \hline
    $M$ & $\big[ \vec{S}_1 \cdot (\vec{S}_2 \times \vec{S}_4) + \vec{S}_2 \cdot (\vec{S}_3 \times \vec{S}_1)$ \\
    & $ \,\,\,\,\,\,\,\,\,\,\, + \, \vec{S}_3 \cdot (\vec{S}_4 \times \vec{S}_2) + \vec{S}_4 \cdot (\vec{S}_1 \times \vec{S}_3)
\big]$ & 3+0.65$\epsilon$ \end{tabular}
\end{ruledtabular}
\caption{\label{tab:enhanced-ops} List of observable in the spin
model that are symmetry-equivalent to the $N^a$ and $M$ fermion
bilinears. For some of these we label the sites around the plaquette
with lower-left corner at $\vec{r}$ by the numbers $1,\dots,4$.
Precisely, $\vec{S}_1 = \vec{S}_{\br}$, $\vec{S}_2 = \vec{S}_{\br +
\bx}$, $\vec{S}_3 = \vec{S}_{\br + \bx + \by}$ and $\vec{S}_4 =
\vec{S}_{\br + \by}$. } \label{bil}\end{table}

Here we have defined the  observable as:
\begin{eqnarray}
\vec{N_A}^i  = & -i\bar{\psi}\mu^i\vec{\sigma}\psi \\
\vec{N_B}  = & -i\bar{\psi}\vec{\sigma}\psi\\
{N_C}^i  = & -i\bar{\psi}\mu^i\psi\\
M = & -i\bar{\psi}\psi
\end{eqnarray}
 We see that $\vec{N_A}^z$ corresponds to Neel vector and
$iN_c^x+N_c^y$ corresponds to VBS order parameter.

\subsection{Discussion}
Now let us examine the trends shown by the exponents calculated in table \ref{bil}. Note
that the scaling dimension of these gauge invariant bilinears are
the same in ASL phase due to the $SU(4)$ symmetry. So for all of them
we have $\Delta=3-1.94\epsilon$. Remarkably the scaling dimension of the VBS order parameter
($N^x_C, N^y_C$) is {\em smaller} at the critical point than it is in the ASL phase.
Thus the VBS fluctuations are {\em enhanced} by the critical $\vec N$ vector fluctuations.
 All other fermion bilinears
decay faster at the critical point as their scaling dimension is increased. This includes the
vector $\vec{N_A}^z$ which is the contribution from the gapless fermions to the
Neel vector. It is at present not clear whether in $d = 2$ the susceptibilities of these
other operators (such as $\vec N^x_A$, etc) will diverge at the critical point (though they apparently do in the ASL
phase). In contrast the VBS susceptibility will presumably diverge at the critical point.
The divergence will be  faster at the critical point
(as say a function of temperature) than in the ASL phase. On the magnetic side the
VBS susceptibility will of course be finite in the ground state. However it
will diverge as the transition is approached and will thus be large if the antiferromagnet is
to be regarded as being close to this critical point. Thus a qualitative conclusion from
our calculations is that in the limit that the Neel state can be usefully regarded as being born out of
the dRVB spin liquid, it will also have enhanced VBS susceptibility.

The diverging VBS susceptibility also provides an interesting way to
define the corelation length $\xi$ in terms of directly measurable
quantities. Consider the VBS corelations in the magnetic side as a
function of length scale. At scales smaller than $\xi$ they will
decay as a power law. However at scales larger a length set by
$\xi$, they will decay exponentially. Thus $\xi$ may be usefully
defined as the correlation length for VBS fluctuations in the
ordered antiferromagnet.

\section{Implications for cuprate theory}
\label{hitcimp}
We now discuss some of
the implications of our results for theories of the cuprates. As we
emphasized in the introduction, a second order transition between a
collinear Neel state and a gapless spin liquid is attractive for a
number of reasons. Here we explore this in some greater detail. Our
thinking on the cuprates is guided by Fig \ref{phase}. We suppose
that increasing magnetic frustration (the parameter $g$) at zero
doping can induce a transition out of the collinear Neel state to a
spin liquid. Theoretically the spin liquid is expected to evolve
rather naturally into a superconductor when it is doped. The real
material starts off in the antiferromagnetic state at zero doping.
The idea is that doping (apart from introducing holes) also has the
effect of increasing $g$. Then we can hope that the intermediate and
long scale physics of the resulting doped superconductor may be
fruitfully described as a doped spin liquid. This is the rationale
behind the spin liquid based approach to the cuprates.

With this point of view in mind let us consider the effects of
doping the dRVB algebraic spin liquid phase discussed in this paper.
Previous papers (for a review see Ref. \onlinecite{lnwrmp}) have shown how a $d$-wave
superconductor with gapless nodal quasiparticles emerges quite
naturally upon doping this spin liquid. Now consider reducing the
doping in the real material. According to Fig \ref{phase} this also
has the effect of reducing the magnetic frustration $g$. This pushes
the ``parent" spin liquid state closer to the transition to
antiferromagnetism. The magnetic response of the parent spin liquid
at wave vector $(\pi, \pi)$ then evolves in the manner shown in Fig
\ref{sc}. Note in particular that if the transition to the magnetic
state is second order then the ``resonance" due to the $\vec N$
fluctuations softens. How does this impact the magnetic response in
the doped superconductor?

The doping of the spin liquid is incorporated theoretically by the
introduction of two species of charged spinless bosons $b_1$ and $b_2$. These bosons also carry gauge charges $+1$ and $-1$ respectively. Superconductivity is achieved when both $b_1$ and $b_2$ condense with equal amplitude.
This route from the dRVB spin liquid to the dSC has two important features. First the gauge charge carried by the bosons
implies that the gapless gauge fluctuations of the spin liquid are quenched in the superconducting state (by the Anderson Higgs mechanism). The spinons evolve naturally into the fermionic quasiparticle excitations of the dSC.
The nodal structure of the spinons is retained - however coupling between the $b$ and $f$ fields moves the nodes of the
quasiparticles away from $(\pi/2, \pi/2)$ by an amount proportional to the doping $x$.

For the magnetic response this has some crucial implications. First
when compared with Fig \ref{sc}, the diverging low frequency
response is killed as it comes entirely due to the gapless gauge
fluctuations of the spin liquid. The resonance due to the triplon
$\vec N$ mode then becomes the most prominent feature in the $(\pi,
\pi)$ response. Further its frequency will soften as the doping is
reduced. Second as the fermionic quasiparticles no longer have nodes
at $(\pi, \pi)$ they only weakly damp out this reonance. Finally the
fermionic quasiparticles will still contribute some background
magnetic response which can now be usefully addressed in a standard
RPA calculation. Such calculations have been reported before in the
literature, and give rise to incommensurate continuum scattering at
frequencies below the resonance that appear to be consistent with
experiment.

Since the original discovery of the neutron resonance peak, there
have been two more or less independent interpretations. One view is
to describe it as a soft mode associated with the magnetism of the
undoped Mott insulator. This view has the advantage that it provides
a natural explanation of the softening of the resonance frequency
with underdoping. The other view has been to simply regard it as a
$S = 1$ collective mode of weakly correlated fermionic
quasiparticles in the superconducting state. The description given
above unifies these two different interpretations. Indeed in the
parent spin liquid the triplon $\vec N$ mode may be viewed as a
particle-hole triplet exciton made out of {\em spinons} - rather
than electrons. This mode appears naturally as a recognizable peak
in the magnetic response upon approaching the AF state. Doping this
spin liquid then leads to a superconductor with gapless fermionic
quasiparticles and a sharp gapped $S = 1$ triplon.

\section{Conclusion}
\label{concl}

In this paper we have revisited the issue of possible second order
phase transitions out of the collinear Neel state into paramagnetic
spin liquid states in two dimensional quantum antiferromagnets. The
particular spin liquid we considered is a $dRVB$ state which has
gapless spin excitations. Correspondingly there are non-trivial
power law corelations in the spin and other quantities. A useful
description is provided in terms of gapless Dirac-like spinons that
are coupled to a fluctuating $U(1)$ gauge field. However there is
possibly no true quasiparticle description of the spectrum. Indeed
this state is in a critical phase that is the two dimensional analog
of the one dimensional spin-$1/2$ chain. In contrast to other
simpler spin liquids which have a spin gap, a direct second order
transition to the collinear Neel state appears to be possible for
such a two dimensional algebraic spin liquid. We developed in some
detail a theory for such a transition. Magnetic long range order was
obtained as a spin density wave transition of spinons. We argued
that gauge fluctuations convert the resulting magnetic state into a
conventional one that is smoothly connected to the usual Neel state.
Thus the spinons disappear from the spectrum in the magnetic state.
The theory for the transition shares a number of similarities with
the deconfined critical points studied recently. Most importantly,
there are two diverging length/time scales as the transition is
approached from the magnetic side. The shorter of the two scales is
associated with the onset of magnetic order from a critical soup of
spinons. The second longer scale is associated with confinement of
the spinons. It is in the intermediate length/time scale regime
({\em i.e.} between the two diverging lengths) that the magnetic
ordering is correctly described as a spin density wave formed out of
spinons. This intermediate scale regime may also be characterized as
a fractionalized antiferromagnet.

We noted several implications of our results for theories of the cuprates that regard them as doped spin liquids.
First it allows us to develop a qualitative picture of the resonance mode seen in neutron experiments.
Our picture unifies the existing descripions as a soft mode associated with the magnetic ordering in the insulator
and as a triplet excition formed from a particle-hole pair of fermionic BCS quasiparticles. Indeed in our description
the resonance is a soft mode of the magnetic ordering that is formed as a particle-hole triplet exciton of {\em spinons}.
This picture is closest to that in Ref. \cite{bmanlee}.

We have shown how magnetism may be incorporated into the spin liquid based approach to the cuprates.
Central to this is the description of magnetism as a spin density wave ordering of spinons. Such a description
has been explored before in a number of publications. As summarized in the first paragraph of this section,
our work clarifies the range of validity of such a description. Indeed should experiments reveal clear signatures
for a fermionic spinon description of the intermediate scale spin physics of the undoped cuprates
then we could take that to be
a signature of proximity to the quantum transition to the dRVB algebraic spin liquid.

\section*{Acknowledgements}
We thank Patrick Lee, M. Hermele, and X.-G. Wen for useful discussions.
This work was supported by NSF Grant No. DMR-0308945. TS also
acknowledges funding from the NEC Corporation, the Alfred P. Sloan
Foundation, and an award from the The Research Corporation.

\appendix
\section{Dirac  Action}\label{Dirac}
Here we show that low energy effective action is described by a
continuum theory of fermionc spinons with Dirac dispersion. As
mentioned before, we need to expand the Hamiltonian close to nodes
at $(\frac{\pi}{2},\frac{\pi}{2})$ and
$(-\frac{\pi}{2},\frac{\pi}{2})$. From now on $(k_x,k_y)$ refer to
deviation from $(\frac{\pi}{2},\frac{\pi}{2})$ or
$(-\frac{\pi}{2},\frac{\pi}{2})$ points. Here we explicitly derive
the Hamiltonian near $(\frac{\pi}{2},\frac{\pi}{2})$. The other node
is similar:
\begin{equation}
H=\left[\begin{array}{ccc} -\frac{4}{g}\vec N.\frac{\vec \sigma}{2} & 2t\ k_+\ -\ 2i\Delta\ k_-\\
 2t\ k_+\ +\ 2i\Delta\ k_- & \frac{4}{g}\vec N.\frac{\vec
\sigma}{2}
\end{array}\right]
\end{equation}
where $k_+=k_x+k_y$ and $k_-=k_x-k_y$. This Hamiltonian could be
written in terms of two by two Pauli matrices which are defined as:
\begin{equation}\nonumber
\tau^x=\left[\begin{array}{ccc}0 & \ 1\\1 &\ 0\end{array}\right]\ \
\  \tau^y=\left[\begin{array}{ccc}0 & -i\\i &0\end{array}\right]\ \
\  \tau^z=\left[\begin{array}{ccc}1 & 0\\0 &-1\end{array}\right]
\end{equation}
Using this notation and also adding the contribution from
$(-\frac{\pi}{2},\frac{\pi}{2})$ node we get the following form for
low energy Hamiltonian:
\begin{equation}\label{stpoint}\begin{split}
H=\sum_{k_+,k_-} & c_1^{\dag}\ (2t\  k_+\tau^x+2\Delta\
k_-\tau^y-\frac{4}{g}\ \tau^z\  \vec N .\frac{\vec \sigma}{2})\  c_1\\
& -c_2^{\dag}\ (2t\ k_-\tau^x+2\Delta\  k_+\tau^y+\frac{4}{g}\
\tau^z\ \vec N .\frac{\vec \sigma}{2})\ c_2
\end{split}
\end{equation}
Here $c_1$ and $c_2$ refer to $(\frac{\pi}{2},\frac{\pi}{2})$ and
$(-\frac{\pi}{2},\frac{\pi}{2})$ nodes, respectively. They are two
component fermionic operators, each component representing one of
the sites in the unit cell. The $\tau^i$ matrices operate in the
space of these two components. Each component has a $SU(2)$ spin
index, where $\sigma$ matrices operates. Now assume $t=\Delta$ (i.e.
ignore the velocity un-isotropy which is irrelevant in
renormaliztion group language) and we rename $k_+$ as $k_y$ and
$k_-$ as $k_x$. Then introduce the new fermionic operators:
\begin{eqnarray}
\psi_1&=&-i\tau^x\ c_1\\
\psi_2&=&e^{i\frac{\pi}{4}\tau^z}\ c_2
\end{eqnarray}
and subsequently:
\begin{equation}
\bar{\psi}_{1,2}=\psi_{1,2}^{\dag}(i\tau^z)
\end{equation}

With these new variables, the Hamiltonian (\ref{stpoint}) takes the
following simple form:
\begin{equation}\begin{split}
H=\sum_{k_x,k_y} & \bar{\psi}_1( k_x\tau^x+k_y\tau^y+J\vec N
.\frac{\vec \sigma}{2})\ \psi_1\\ &+ \bar{\psi}_2(
k_x\tau^x+k_y\tau^y-J\vec N .\frac{\vec \sigma}{2})\ \psi_2
\end{split}
\end{equation}
Then using $\mu$ matrices that operates in the space, made by
presence of two different nodes, we can put but nodes contribution
($\psi_1$ and $\psi_2$) as a single vector, $\psi$. Then the
Hamiltonian takes the form:
\begin{equation}
H=\sum_{k_x,k_y}  \bar{\psi}( k_x\tau^x+k_y\tau^y+i J\mu^z\ \vec N
.\frac{\vec \sigma}{2})\ \psi
\end{equation}
 This form in continuum limit and in real space leads to action given in \ref{maction}.
\section{Random Phase Calculation}\label{rpa}
We start with the partition function with an external magnetic
field $h$ that couples to the Neel vector:
\begin{eqnarray}
Z&=&\int [D\psi][D\bar{\psi}][Da][DN]\  e^{-S}\\
S&=&S_{N}+S_{\psi,a}+S_{mixing}+S_h\\
S_{N}&=&\int d^2x d\tau \{\frac{1}{2}(\partial_{\mu}
N)^2+\frac{r}{2} N^2\}\\
S_{\psi,a}&=& \int d^2x d\tau
\bar{\psi}[-i\gamma^{\mu}(\partial_{\mu}+i e a_{\mu})]\psi\\
S_{mixing}&=& \int d^2x d\tau \ \
i\bar{\psi}\vec\sigma\mu^z\psi.(\lambda\vec N)\\
S_h&=&\int d^2x d\tau \vec{h}.(a \ i\bar{\psi}\vec\sigma\mu^z\psi+ b
\vec{N})
\end{eqnarray}
Here $a$ and $b$ are non-universal constants that depend on details of the microscopic physics.

 Now by expanding the the terms which contain $i\bar{\psi}\mu^z\vec\sigma\psi$ (one coming
 from $S_{mixing}$ and one from coupling to $\vec h$ field.)
    up to quadratic order, and
performing the integral over $\psi$ and $a$ fields we get the
following form for the partition function:
\begin{equation}\label{eff}
Z=\int [DN] e^{-S_N-b \int d^2x d\tau \vec{h}.\vec{N}}\ \int d^2 q
d\omega \frac{\chi(q,\omega)}{2}|\vec{h}+\lambda\vec{N}|^2
\end{equation}
Where $\chi(q,\omega)$ is the susceptibility of
$i\bar{\psi}\vec\sigma\mu^z\psi$ operator calculated using the
action $S_{\psi,a}$. We can rewrite the
$\frac{\chi(q,\omega)}{2}|\vec{h}+\lambda\vec{N}|^2$ up to quadratic
order in $N$ as an exponential. Putting this form in \ref{eff} we
get the following for  the partition function:
\begin{eqnarray}
Z&=&\int [DN] e^{-S_N^{eff}}\\ \nonumber S_N^{eff}&=&S_N+ \int d^2q
d\omega \{b\ \vec{h_q}.\vec{N_{-q}}- \frac{\chi(q,\omega)}{2}|a\
\vec{h_q}+
\lambda\vec{N_q}|^2\}\\
 \nonumber &=&\int d^2q d\omega \{\frac{\chi_N(\omega,q)^{-1}}{2}
|\vec{N}_q|^2+\vec{h_q}.\vec{N_{-q}}\\
& & -\frac{\chi(q,\omega)}{2}|\vec{h}+\lambda\vec{N}|^2\}
\end{eqnarray}
Now to get the quadratic action for $\vec h$ field we should
integrate out $\vec N$ field, which is a simple Gaussian integral
now. This leads to the following value for effective susceptibility:
\begin{equation}
\chi_{eff}=\frac{a^2 \ \chi +b^2 \ \chi_N-2\lambda a\ b\ \chi\
\chi_N}{1-\lambda^2\chi\ \chi_N}
\end{equation}
 You can see that as one expects, for $\lambda=0$, this reduces to
 some of susceptibilities for $\vec N$ and $i\bar{\psi}\mu^z
 \frac{\vec\sigma}{2}\psi$ fields. Also note that, this is valid only
 for the frequencies where $\lambda^2\chi\ \chi_N$ is not of order
 one ({\em i.e.} it is not valid for small frequencies).
 Now by analytic continuation to real frequencies and taking the imaginary part,
 we get the spectral functions plotted in section \ref{sp}.
\section{Feynman diagrams}\label{feynman}
\subsection{Spin-Spin correlation in mean-field theory}
For spin-spin correlation we need to calculate the following diagram
which consists of two fermionic propagators. The fermionic
propagator in mean field is derived from the action \ref{maction}:
\begin{equation}
\feyn{ f\vertexlabel_{p}f }\ = \frac{1}{\pslash-i\lambda
\vec{N}.\vec{\sigma}}
\end{equation}
\begin{figure}[htp]
\includegraphics[height=1.3cm]{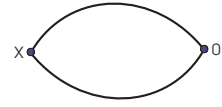}
\caption{Spin-Spin correlation in mean field}
\label{spsp}\end{figure}
 This diagram, in the real space, corresponds to
the following integral:
\begin{equation}\begin{split}\langle S_i S_j \rangle=-\delta_{ij}
\int & \frac {d^3 p}{(2\pi)^3} \frac{d^3 q}{(2\pi)^3}\\
& e^{-i\vec{p}.\vec{r}}\frac{tr[(\pslash+\qslash-i\lambda
\vec{N}.\vec{\sigma})(\qslash-i\lambda\vec{N}\vec{\sigma})]}{((p+q)^2+\lambda^2
N^2)(q^2+\lambda^2N^2)}
\end{split}
\end{equation}
After integrating over $q$ this gives:
\begin{equation}\begin{split}
\langle S_i S_j \rangle \propto -\frac{\delta_{ij}}{r}\int p \ dp &
\ sin(pr)[4\lambda N +\\ & \frac{2}{p}\arctan(\frac{p}{2\lambda
N})(4\lambda^2N^2+p^2)]
\end{split}
\end{equation}
The first  term is proportional to $\delta(r)$. Since we are
interested in the case where $r\neq0$, we can ignore that term. The
second term gives:
\begin{equation}
\frac{e^{-2\lambda N r}}{r^4}(1+2\lambda N r)
\end{equation}
We already saw that $N$ goes to zero as $g$ approaches $g_c$ like
$g_c-g$. So if we define correlation length as
$\xi=\frac{1}{2\lambda N}$ we get the equations \ref{corm} and
\ref{corlm}.

\subsection{Beyond mean field: $\epsilon$-expansion}

 Now to do the $\epsilon$-expansion we need to study the action given in \ref{action}. We can see
that there are three different types of vertices present in the
theory and eleven different one loop diagrams which cause field
renormalizations, vertex corrections and mass renormalization. we
use the following diagrammatic representation for the propagators of
the fields, in the theory:
\begin{equation}\begin{split}
\feyn{\vertexlabel^a h\vertexlabel_{p}h \vertexlabel^b}\ & =
\frac{\delta^{ab}}{p^2+r}\\
\feyn{ f\vertexlabel_{p}f }\ & = \frac{1}{\pslash}\\
\feyn{\vertexlabel^{\mu} g\vertexlabel_{p}g \vertexlabel^{\nu}}\ & =
\frac{\delta^{\mu \nu}}{p^2}
\end{split}
\end{equation}
Here we have used Feynman gauge for the gauge field propagator, and
Euclidian metric. Now we start to calculat the one loop diagrams.

\subsubsection{Fermion self energy} There are two, one loop
diagrams which will generated fermions self energy.
\begin{figure}[htp]
\centering
\begin{center}
\subfigure[]{\label{sea}
\includegraphics[height=1.5cm]{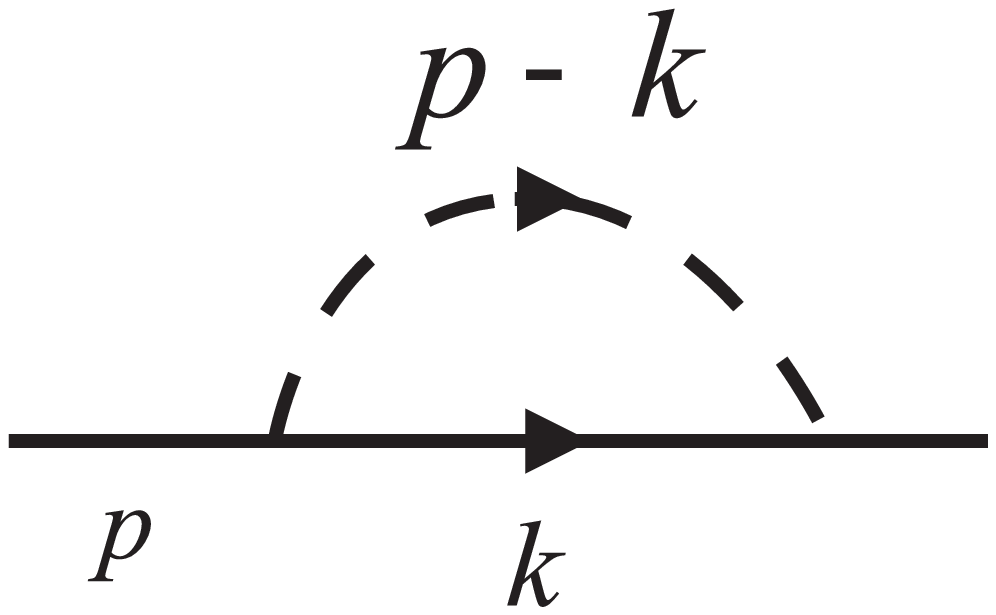} \vspace{0.1in}
\hspace{.3in}} \subfigure[]{
\includegraphics[height=1.5cm]{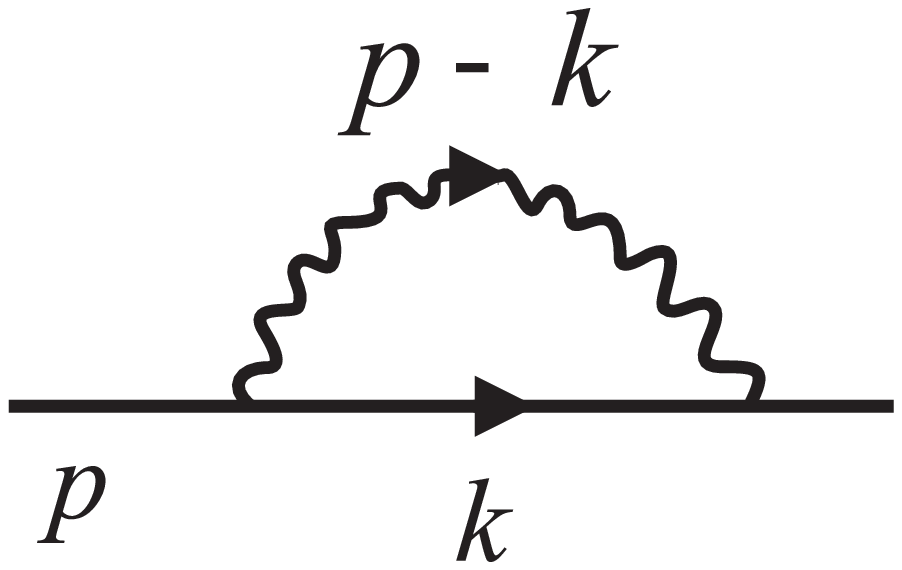}
\vspace{0.1in} \label{seb}} \caption{fermion self energy}
\end{center}
\end{figure}
Diagram \ref{sea}  represent $N$ field contribution to fermion self
energy:
\begin{equation}\label{esa}\begin{split}
\Sigma^{\psi}_{N} \pslash&= (i\lambda)^2\int \frac{d^d
k}{(2\pi)^d}\sigma^i\frac{\delta^{ij}\ \kslash}{k^2(k+p)^2}\sigma^j\\
&=-\frac{3}{(4\pi)^2}\frac{\lambda^2}{\epsilon}\pslash
\end{split}
\end{equation}
 Diagram \ref{seb} represents gauge field contribution to fermion
self energy:
\begin{equation}\label{esb}\begin{split}
\Sigma^{\psi}_{a} \pslash&= e^2\int \frac{d^d
k}{(2\pi)^d}\gamma^{\mu}\frac{\delta^{\mu\nu}\ \kslash}{k^2(k+p)^2}\gamma^{\nu}\\
&=-\frac{1}{(4\pi)^2}\frac{e^2}{\epsilon}\pslash
\end{split}
\end{equation}
\subsubsection{N Field Self Energy} Since $N$ field is coupled to
fermion field only, there is just one diagram contributing to its
self energy.
\begin{figure}[htp]
\centering
\includegraphics[height=1.3cm]{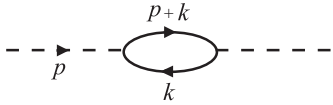}\vspace{0.1in}
\caption{N field self energy}
\end{figure}
\newline This has the following contribution:
\begin{equation}\begin{split}
\Sigma^{N}_{\psi} p^2\delta^{ij}&=(-1) (i\lambda)^2\int \frac{d^d
k}{(2\pi)^d}\frac{tr(\sigma^i\sigma^j)tr(\kslash(\kslash+\pslash))}{k^2(k+p)^2}\\
&=-\frac{8}{(4\pi)^2}\frac{\lambda^2}{\epsilon}p^2\delta^{ij}
\end{split}
\end{equation}
Here the minus sign appears because of presence of a fermionic loop
in the diagram \cite{ps}.
\subsubsection{Gauge field self energy} Gauge field
is only coupled to fermion field as well, so there is only one
diagram generates gauge field self energy:
\begin{figure}[htp]
\centering
\includegraphics[height=1.3cm]{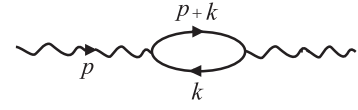}\vspace{0.1in}
\caption{Gauge Field Self Energy}
\end{figure}

This diagram has the following contribution:
\begin{equation}\begin{split}
\Sigma^{a}_{\psi}[p^{\mu}p^{\nu}-p^2\delta^{\mu\nu}] &=(-1) e^2\int
\frac{d^d
k}{(2\pi)^d}\frac{tr(\gamma^{\mu}\kslash\gamma^{\nu}(\kslash+\pslash))}{k^2(k+p)^2}\\
&=-\frac{16}{(4\pi)^2}\frac{e^2}{3\epsilon}p^2[\delta^{\mu
\nu}-\frac{p^{\mu}p^{\nu}}{p^2}]
\end{split}
\end{equation}
Note that in this relation the term proportional to $p^{\mu}p^{\nu}$
dose not contribute to physical observables (S-matrix elements).
This is guarantied by Ward identity\cite{ps}.
\subsubsection{$u((\vec N)^2)^2$ vertex correction} There are two diagrams contributing
to this vertex renormalization at one loop level.
\begin{figure}[htp]
\centering
\begin{center}
\subfigure[]{\label{vua}
\includegraphics[height=1.9cm]{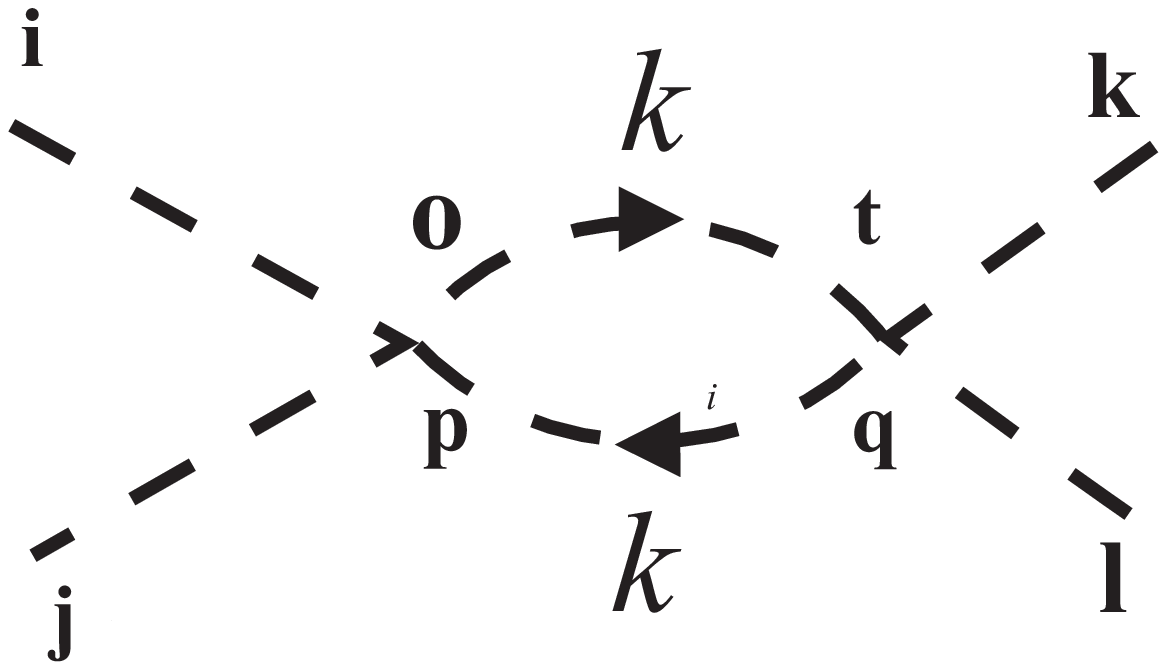} \vspace{0.1in}
\hspace{.3in}} \subfigure[]{
\includegraphics[height=1.9cm]{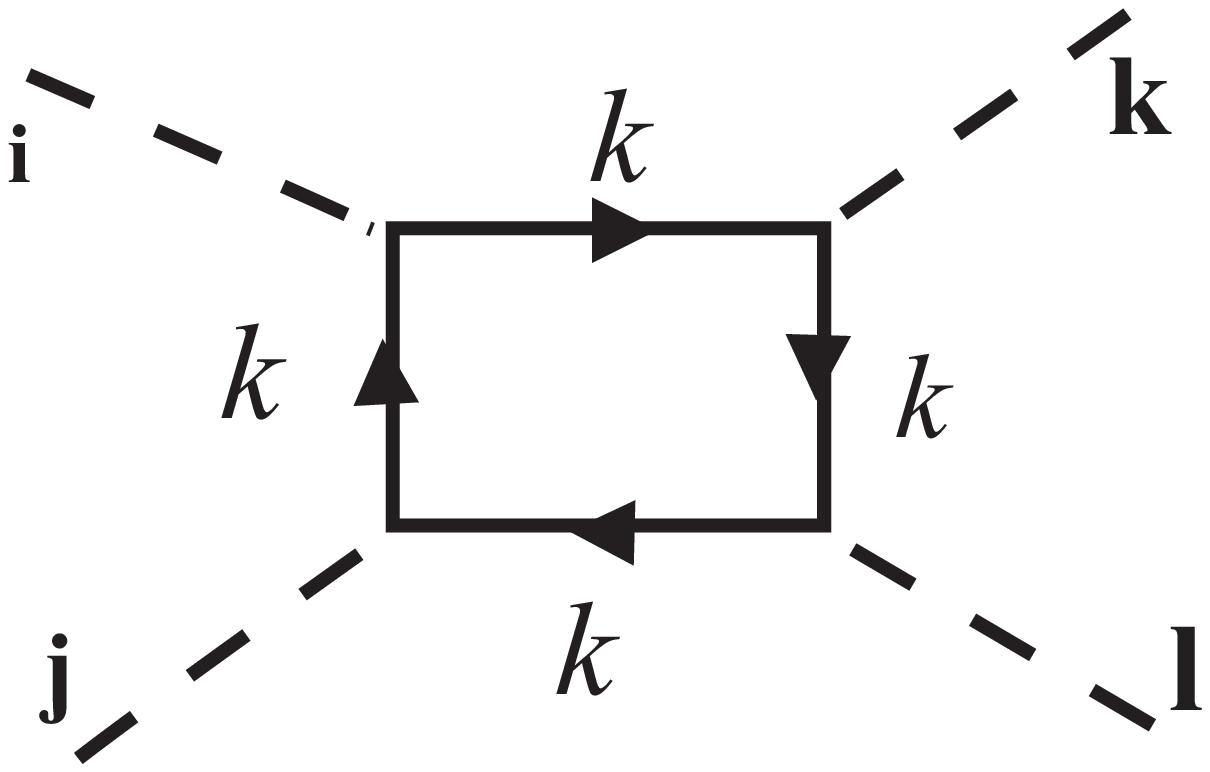}
\vspace{0.1in} \label{vub}} \caption{u renormalization}
\end{center}
\end{figure}
Note that  up to one loop order, it is enough  to calculate vertex
corrections with zero external momentum. Diagram \ref{vua} gives:
\begin{equation}\begin{split}
\Delta^{u}_N
(\delta^{ij}\delta^{kl}+\delta^{ik}&\delta^{jl}+\delta^{il}\delta^{jk})\\
=-\frac{u^2}{6}(\delta^{ij}\delta^{op}&+\delta^{io}\delta^{jp}+\delta^{ip}\delta^{jo})\int\frac{d^d
k}{(2\pi)^2}\frac{\delta^{oq}\delta^{pt}}{(k^2+r)^2}
\\&(\delta^{qt}\delta^{kl}+\delta^{qk}\delta^{tl}+\delta^{ql}\delta^{tk})\\
=-\frac{11}{(4\pi)^2}\frac{u^2}{3\epsilon}(&\delta^{ij}\delta^{kl}+\delta^{ik}\delta^{jl}+\delta^{il}\delta^{jk})
\end{split}
\end{equation}
The contribution of diagram \ref{vub} is similarly calculated:
\begin{equation}\begin{split}
\Delta^{u}_{\psi}
&(\delta^{ij}\delta^{kl}+\delta^{ik}\delta^{jl}+\delta^{il}\delta^{jk})\\
&=-3(-1)(i\lambda)^4\int\frac{d^d k}{(2\pi)^d}\frac{tr(\kslash^4)}{k^8}tr(\sigma^i\sigma^j\sigma^k\sigma^l)\\
&=\frac{96}{(4\pi)^2}\frac{\lambda^4}{\epsilon}(\delta^{ij}\delta^{kl}+\delta^{ik}\delta^{jl}+\delta^{il}\delta^{jk})
\end{split}
\end{equation}
Again one minus sign appears because of the fermionic loop
\cite{ps}.
\subsubsection{N-$\psi$ vertex correction}\label{vort}
Again, there are two different diagrams associated with this
correction at one-loop level:
\begin{figure}[htp]
\centering
\begin{center}
\subfigure[]{\label{vla}
\includegraphics[height=1.6cm]{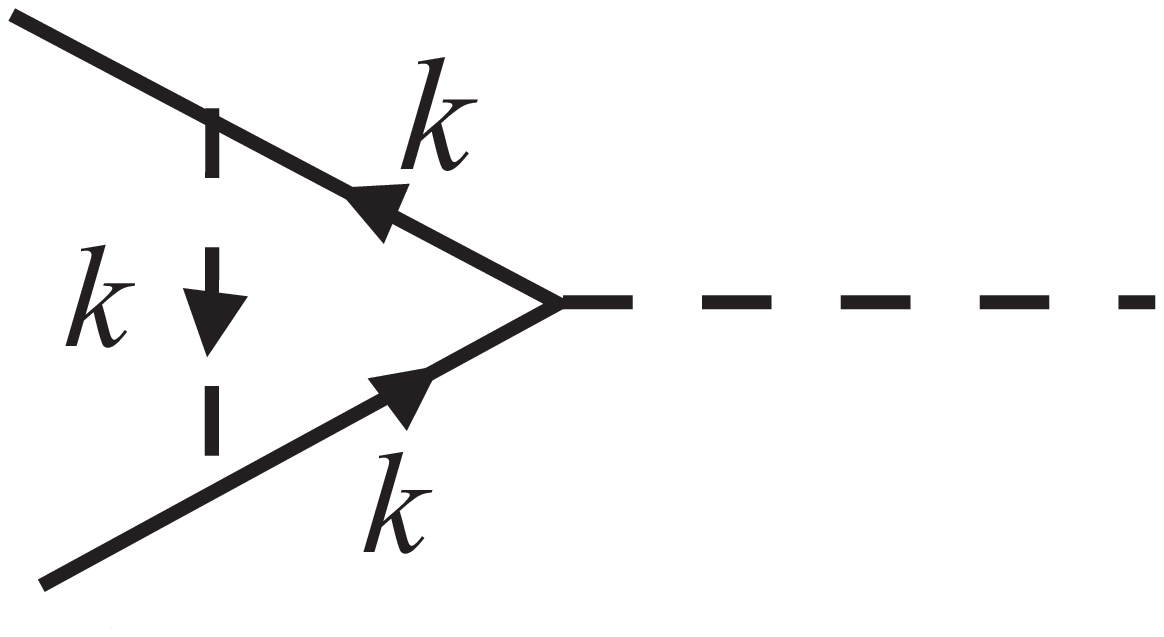} \vspace{0.1in}
\hspace{.3in}} \subfigure[]{
\includegraphics[height=1.6cm]{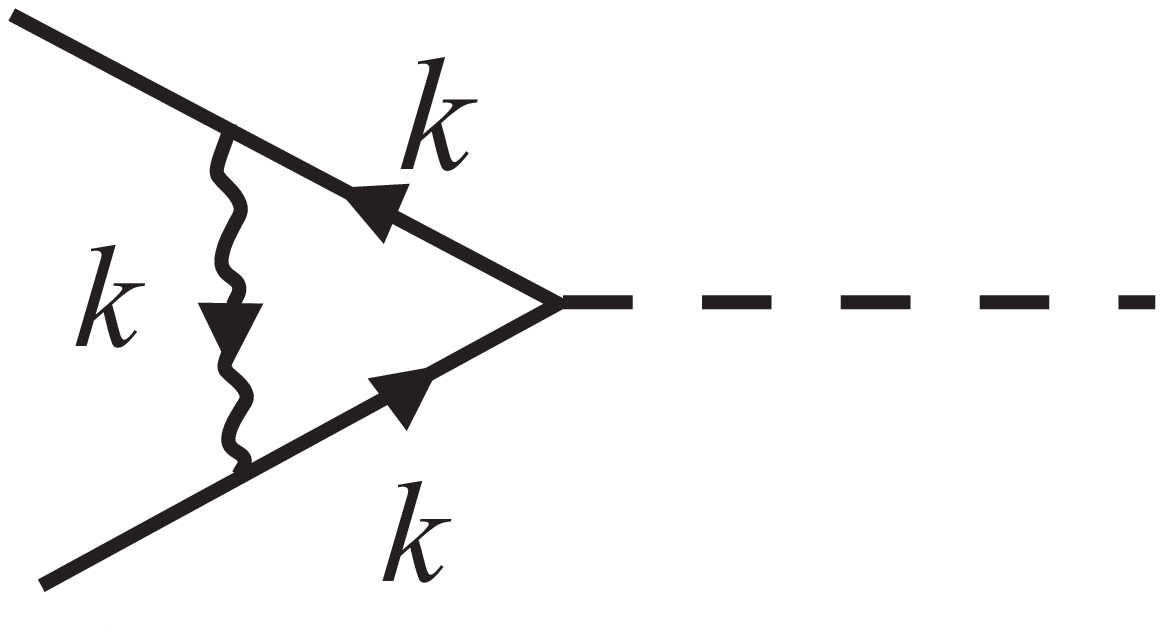}
\vspace{0.1in} \label{vlb}} \caption{$\lambda$ renormalization}
\end{center}
\end{figure}
\newline As before we set the external momentums to zero. Then the
contribution of diagram \ref{vla} is:
\begin{equation}\begin{split}
i\Delta^{\lambda}_N \sigma^k=& (i\lambda)^3\int\frac{d^d
k}{(2\pi)^d}\frac{\kslash}{k^2}\sigma^i\frac{\ \delta^{ij}\ \sigma^k\ }{k^2+r}\ \sigma^j\frac{\kslash}{k^2}\\
=&\ i\frac{2}{(4\pi)^2}\frac{\lambda^3}{\epsilon}\sigma^k
\end{split}
\end{equation}
and diagram \ref{vlb} gives:
\begin{equation}\begin{split}
i\Delta^{\lambda}_a=& i\lambda\  e^2\int\frac{d^d
k}{(2\pi)^d}\frac{\kslash}{k^2}\gamma^{\mu}\frac{\delta^{\mu\nu}}{k^2+r}\gamma^{\nu}\frac{\kslash}{k^2}\\
=&i\frac{6}{(4\pi)^2}\frac{\lambda\ e^2}{\epsilon}
\end{split}
\end{equation}
\subsubsection{a-$\psi$ vertex correction} Similar to \ref{vort}
there are two diagrams:
\begin{figure}[htp]
\centering
\begin{center}
\subfigure[]{\label{vga}
\includegraphics[height=1.6cm]{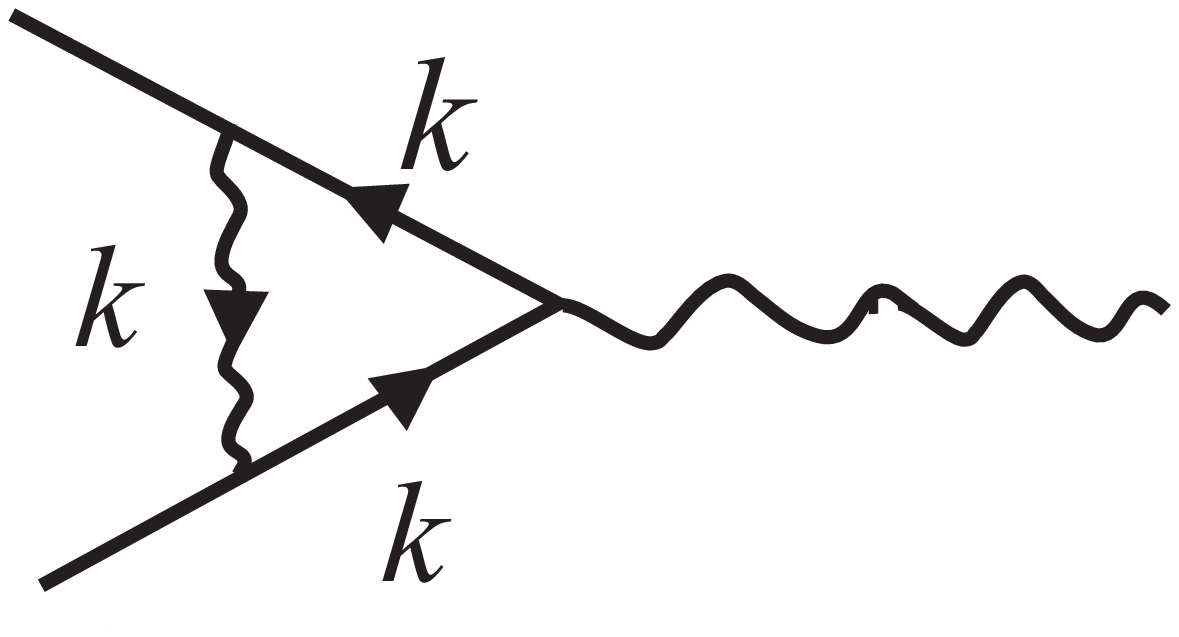} \vspace{0.1in}
\hspace{.3in}} \subfigure[]{
\includegraphics[height=1.6cm]{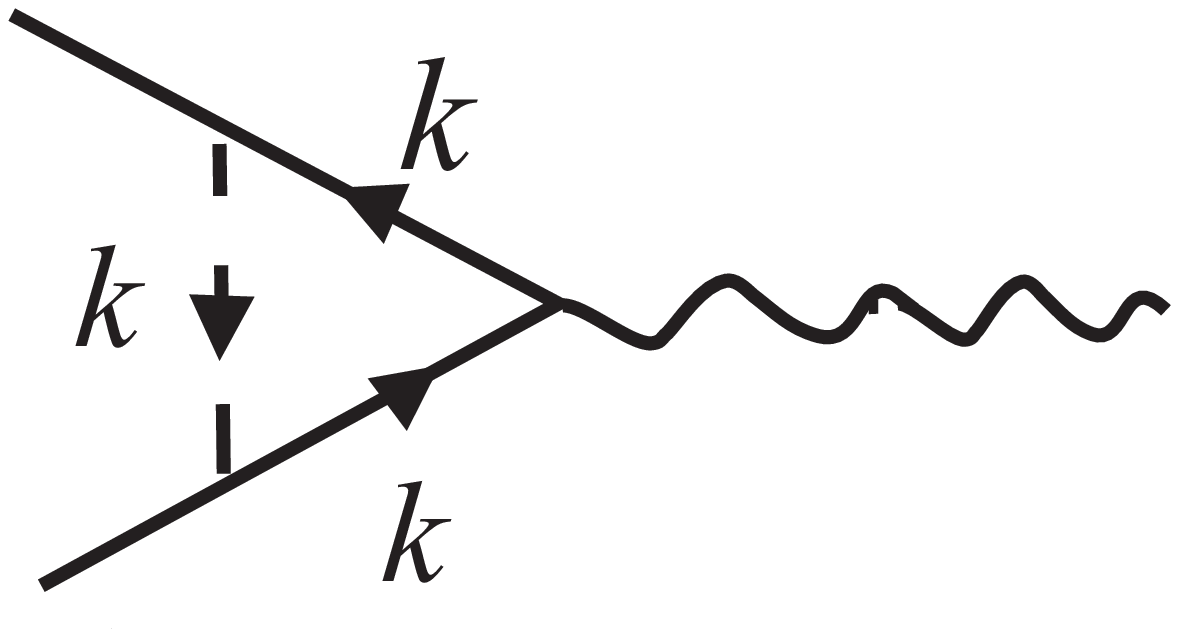}
\vspace{0.1in} \label{vgb}} \caption{e renormalization}
\end{center}
\end{figure}
\newline The first diagram (\ref{vga}) gives:
\begin{equation}\label{asia}\begin{split}
\Delta^{e}_a\ \gamma^{\alpha}=&   e^3\int\frac{d^d
k}{(2\pi)^d}\gamma^{\mu}\frac{\kslash}{k^2}\frac{\gamma^{\alpha}\delta^{\mu\nu}}{k^2}\frac{\kslash}{k^2}\gamma^{\nu}\\
=&\frac{1}{(4\pi)^2}\frac{ e^3}{\epsilon} \gamma^{\alpha}
\end{split}
\end{equation}
and \ref{vgb} gives:
\begin{equation}\label{asib}\begin{split}
\Delta^{e}_{N}\ \gamma^{\alpha}=&   e(i\lambda)^2\int\frac{d^d
k}{(2\pi)^d}\sigma^i\frac{\kslash}{k^2}\frac{\gamma^{\alpha}\delta^{ij}}{k^2+r}\frac{\kslash}{k^2}\sigma^{j}\\
=&\frac{3}{(4\pi)^2}\frac{ e^3}{\epsilon} \gamma^{\alpha}
\end{split}
\end{equation}
Note that \ref{asia} is minus the contribution of \ref{esb} and
\ref{asib} has the minus contribution of \ref{esa}. This, in fact,
should be the case to keep gauge invariance of the theory.
\subsubsection{Mass renormalization} The last one loop diagram, we
consider, generates mass renormalization:
\begin{figure}[htp]
\centering
\includegraphics[height=1.8cm]{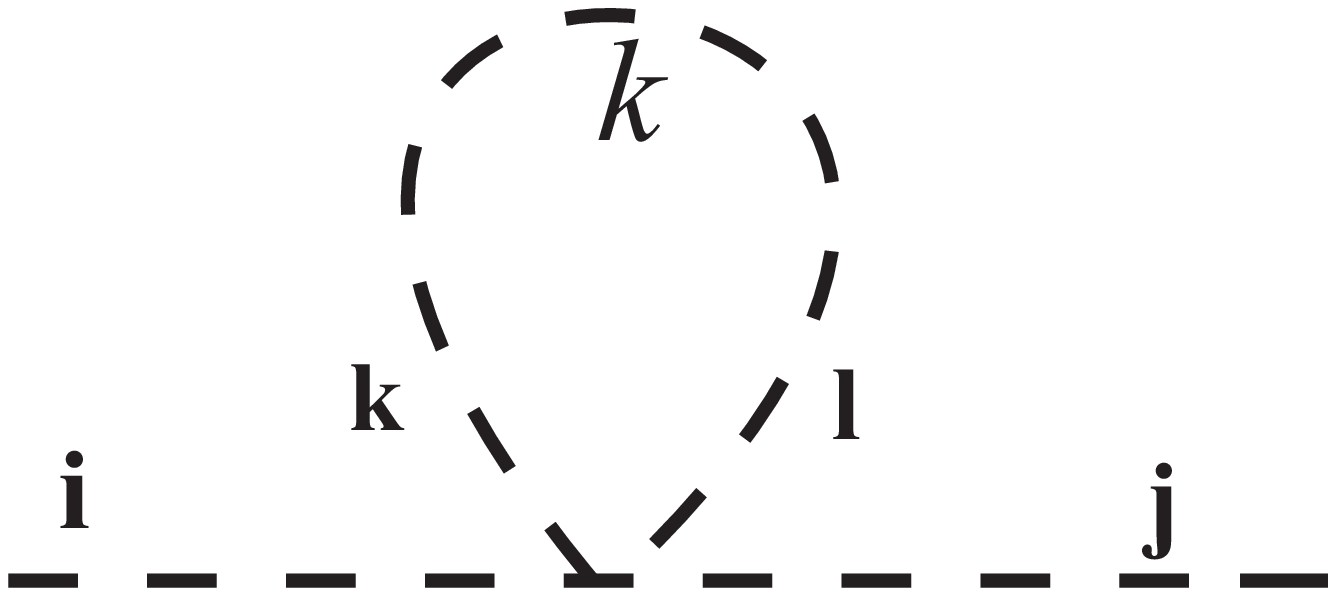}\vspace{0.1in}
\end{figure}
\begin{equation}\begin{split}
\Delta_r\
\delta^{ij}&=\frac{u}{6}(\delta^{ij}\delta^{kl}+\delta^{ik}\delta^{jl}+\delta^{il}\delta^{jk})\int\frac{d^d
k}{(2\pi)^d}\frac{\delta^{lk}}{k^2+r}\\
&=-\frac{5}{(4\pi)^2}\frac{u}{3\epsilon}\ \delta^{ij}
\end{split}
\end{equation}
\subsection{Renormalization conditions}
Now with these in hand, we can proceed using minimal subtraction
scheme. Introducing a mass scale $m \propto \sqrt{r}$ we can write
the following set of renormalization conditions:
\begin{eqnarray}
\nonumber\pslash\
(Z_{\psi}-(\Sigma^{\psi}_N+\Sigma^{\psi}_a))&=&finite\
O(2\  loops)\\
\nonumber m^{-\frac{\epsilon}{2}}e_0
Z_{\psi}\sqrt{Z_a}+(\Delta^{e}_{N}+\Delta^{e}_{a})&=&finite\ O(2\
loops)\\ \nonumber Z_N(p^2+r)+(r\Delta_r - \Sigma^N_{\psi}\
p^2)&=&finite\  O(2\ loops) \\
\nonumber m^{-\frac{\epsilon}{2}}\lambda_0
Z_{\psi}\sqrt{Z_N}+(\Delta^{\lambda}_{N}+\Delta^{\lambda}_{a})&=&finite\
O(2\  loops)\\
\nonumber p^2(Z_a-\Sigma^a_{\psi})&=& finite\ O(2\ loops)
\\
\nonumber m^{-\epsilon}u_0 Z^2_N+ (\Delta^u_N+\Delta^u_{\lambda})&=&
finite\ O(2\ loops)
\end{eqnarray}
Here $e_0$, $\lambda_0$ and $u_0$ are bare coupling constants and so
does not flow with mass scale. These relations give field
renormalization coefficients directly, since the divergence part of
self energy diagrams should cancel out with these field
renormalization coefficients.
\begin{eqnarray}
Z_{\psi}&=&1+\Sigma_{\psi}^N+\Sigma_{\psi}^a\\
Z_N&=&1+\Sigma_{N}^{\psi}\\
Z_a&=&1+\Sigma_a^{\psi}
\end{eqnarray}

 Putting them back in renoemalization condition equations, and letting the mass scale flow \cite{ps}, we get the equations
given in section \ref{eps}.
\subsection{Renormalization of bilinear operatores}
\begin{figure}[htp]
\centering
\begin{center}
\subfigure[]{\label{biln}
\includegraphics[height=1.6cm]{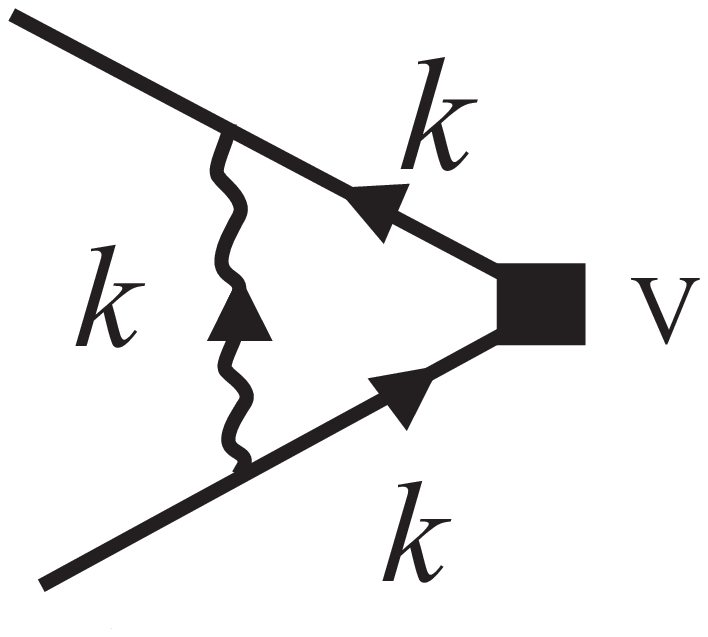} \vspace{0.1in}
\hspace{.3in}} \subfigure[]{
\includegraphics[height=1.6cm]{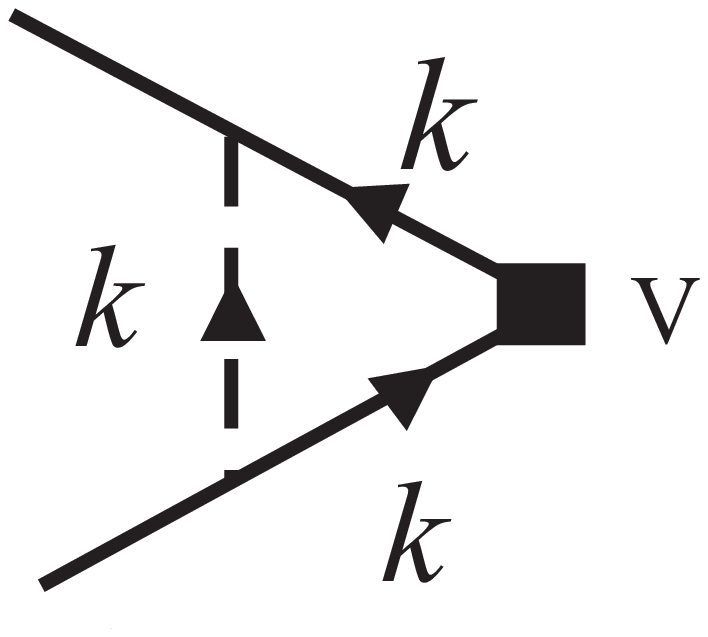}
\vspace{0.1in} \label{billa}} \caption{bilinear operators scaling
dimension}
\end{center}
\end{figure}

 Here we introduce a new term in the Lagrangian with the general
form:
\begin{equation}
v\ \bar{\psi}\cal O \psi,
\end{equation}
where $\cal O$ is the combination of $\mu$ and $\sigma$ matrices
which are given in table \ref{bil}. There are two new diagrams
corresponding to this new vertex (\ref{biln},\ref{billa})
 Diagram \ref{biln} gives:
\begin{equation}\begin{split}
\Delta_{\cal O}^N {\cal O} =&(i\lambda)^2\int \frac{d^d k}{(2\pi)^d}
\sigma^i \mu^z \frac{\kslash}{k^2}{\cal O} \frac{\kslash}{k^2}
\frac{\delta^{ij}}{k^2+r}\mu^z \sigma^j\\ =&-\lambda^2\frac{2
}{(4\pi)^2\ \epsilon}\ A_{\cal O} {\cal O},\ \ A_{\cal O}{\cal
O}=\sigma^i \mu^z {\cal O} \mu^z \sigma^i
\end{split}
\end{equation}
and diagram \ref{billa} gives:
\begin{equation}\begin{split}
\Delta_{\cal O}^a {\cal O} =& e^2\int \frac{d^d k}{(2\pi)^d}
\gamma^{\mu} \frac{\kslash}{k^2}{\cal O} \frac{\kslash}{k^2}
\frac{\delta^{\mu \nu}}{k^2}\gamma^{\nu}\\ =&\ e^2\frac{6
}{(4\pi)^2\ \epsilon}\ {\cal O}
\end{split}
\end{equation}
 Whit these in hand we can get the scaling dimension of $v$:
\begin{equation}\begin{split}
\Delta_{v}=1+\delta_v & =[1+(\Delta_{\cal
O}^N+\Sigma_{\psi}^N)\lambda^2+(\Delta_{\cal
O}^a+\Sigma_{\psi}^a)e^2]v\\
& =[1+(-2A_{\cal
O}-3)\frac{\lambda^2}{(4\pi)^2}+5\frac{e^2}{(4\pi)^2}]v
\end{split}
\end{equation}
This gives the scaling dimension of $v$ ($1+\delta_v$). Now to get
the scaling dimension of $\cal O$ ($\Delta_{\cal O}$), note:
\begin{equation}
\Delta_{\cal O}=D-(1+\delta_v)=3-\epsilon-\delta_v
\end{equation}
 Using the  fixed point values for $\lambda$ and $e$, we get the
scaling dimensions mentioned in table \ref{bil}.
\subsection{Velocity anisotropy}\label{velocity}
 In this section we assume a small velocity anisotropy and treat it
 as a perturbation to our $QED_3$ theory. Following the notation used
 in Ref. \onlinecite{su4}, this anisotropy is presented by:
 \begin{equation}
 K_a=-i\delta\bar{\psi}\mu^z\hat{\gamma^{\mu}}(\partial_{\mu}+ie\ a_{\mu})\psi
 \end{equation}
\begin{figure}[htp]
\centering
\begin{center}
\subfigure[]{\label{ani1}
\includegraphics[height=.8cm]{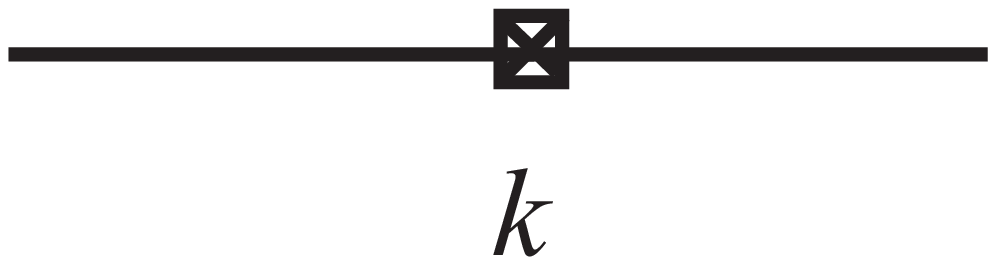} \vspace{0.1in}
\hspace{.3in}} \subfigure[]{
\includegraphics[height=1.3cm]{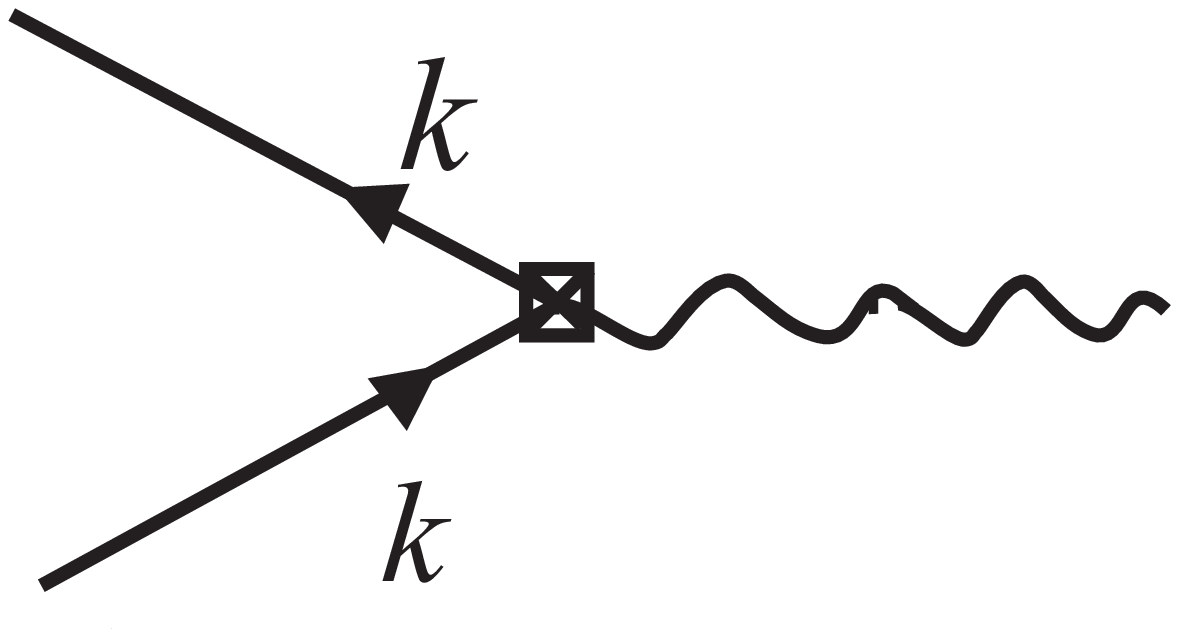}
\vspace{0.1in} \label{ani2}} \caption{Velocity anisotropy vertexes}
\end{center}
\end{figure}

 Here, $\delta$ is the small perturbation parameter (we are
 essential  interested in its behavior under renormalization) and
 $\hat{\gamma^{\mu}}=\gamma^x\delta_{x,\mu}-\gamma^y\delta_{y,\mu}$.
 There are two vertexes associated with this perturbation. One is
 correction in fermionic kinetic energy presented in figure \ref{ani1} and correction
 to the $a-\psi$ vertex presented in figure \ref{ani2}:
\begin{eqnarray}
&-\delta \mu^z \hat{\kslash}
\\&-\delta \mu^z \hat{\gamma^\mu}
\end{eqnarray}

So there are there three different type of one-loop diagrams, in
addition to field renormalization factors calculated before
contributing renormalization of the anisotropy term in fermionic
kinetic energy ({\em i.e.} $\delta$). Diagram \ref{anio1} gives:
\begin{figure}[htp]
\centering
\begin{center}
\subfigure[]{\label{anio1}
\includegraphics[height=1.6cm]{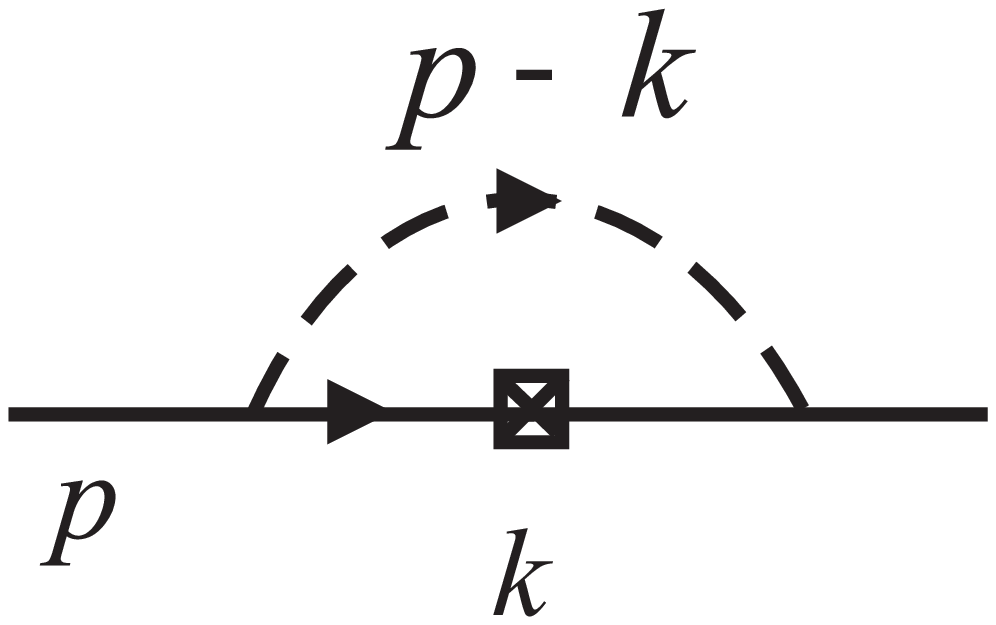} \vspace{0.1in}} \subfigure[]{
\includegraphics[height=1.6cm]{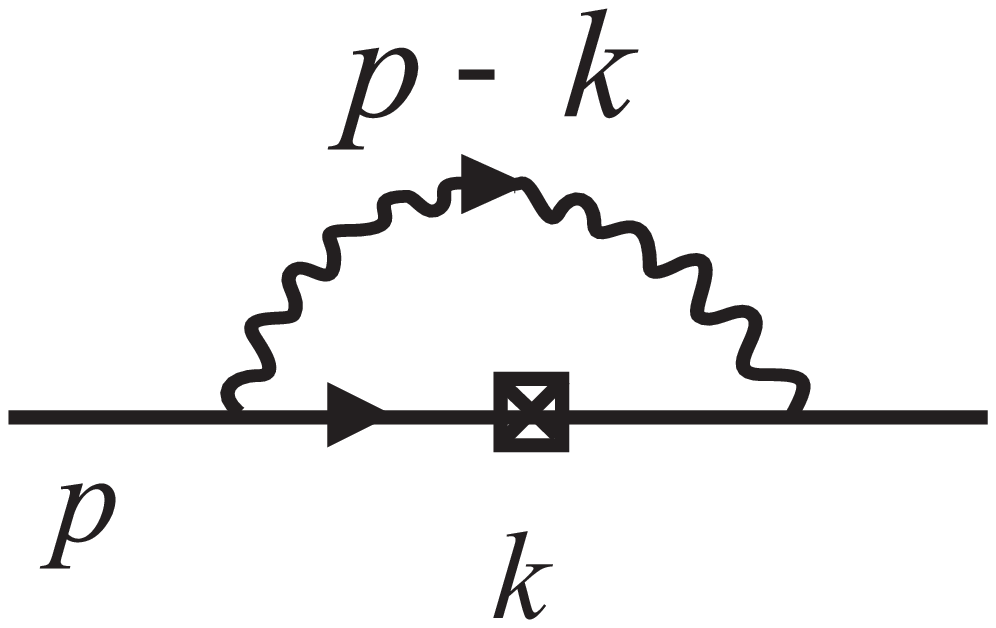}
\vspace{0.1in} \label{anio2}} \subfigure[]{\label{anio3}
\includegraphics[height=1.6cm]{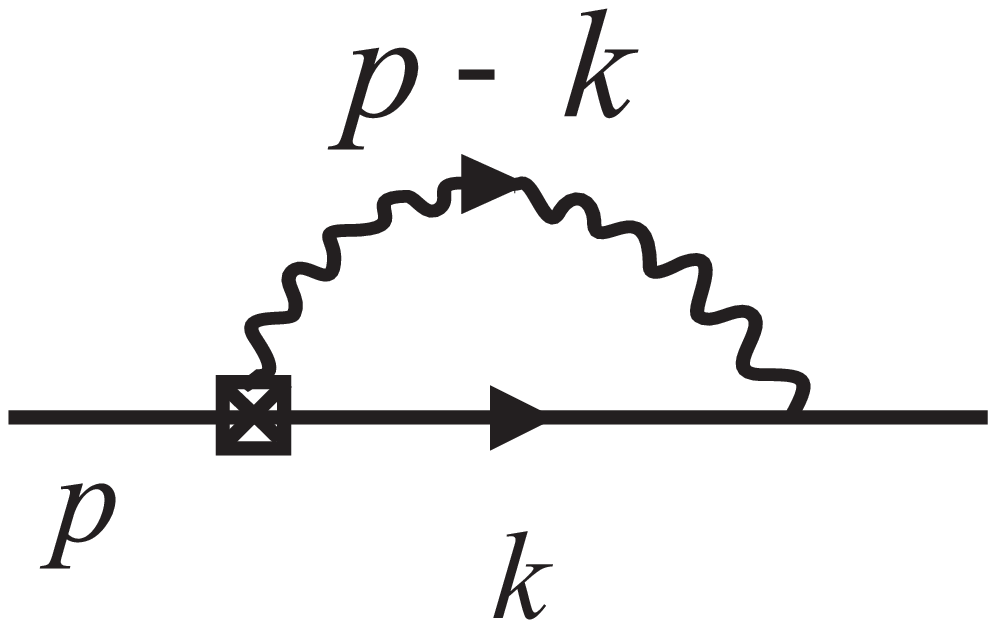} \vspace{0.1in}
} \caption{Velocity anisotropy one-loop renormalization}
\end{center}
\end{figure}
\begin{equation}\begin{split}
-\delta\ \Sigma_{\delta}^N \hat{\pslash}&=-\delta \
(i\lambda)^2\int\frac{d^d k}{(2\pi)^d}\sigma^i
\frac{1}{\kslash}\hat{\kslash}\frac{1}{\kslash}
\frac{\delta^{ij}}{(p-k)^2+r}\sigma^j\\ &=
-\frac{\delta}{(4\pi)^2}\frac{\lambda^2}{\epsilon}\hat{\pslash}
\end{split}\end{equation}
Diagram \ref{anio2} similarly gives:
\begin{equation}\begin{split}
-\delta\ \Sigma_{\delta}^a \delta\hat{\pslash}&=-\delta \
e^2\int\frac{d^d k}{(2\pi)^d}\gamma^{\mu}
\frac{1}{\kslash}\hat{\kslash}\frac{1}{\kslash} \frac{\delta^{\mu
\nu}}{(p-k)^2}\gamma^{\nu}\\ &=
-\frac{\delta}{3(4\pi)^2}\frac{e^2}{3\epsilon}\hat{\pslash}
\end{split}\end{equation}
Diagram \ref{anio3} and the similar one with the correction on the
right vertex give:
\begin{equation}
\begin{split}
-\delta\ \Delta_{\delta}^a \hat{\pslash}&=\delta\ e^2\int\frac{d^d
k}{(2\pi)^d}\hat{\gamma}^{\mu} \frac{1}{\kslash}\frac{\delta^{\mu
\nu}}{(p-k)^2}\gamma^{\nu}\\ &=
\frac{\delta}{(4\pi)^2}\frac{2e^2}{\epsilon}\hat{\pslash}
\end{split}
\end{equation}
Now with these in and also the field renormalziation coefficients
calculated perviously we can get the RG flow for $\delta$:
\begin{equation}\begin{split}
\beta_{\delta}&= \delta\ (2\ \Delta^a_{\delta}+\Sigma^{N}_{\delta}+\Sigma^{a}_{\delta}+\Sigma^{\psi}_N+\Sigma^{\psi}_a)\\
&= -\delta\ (\frac{14}{3(4\pi)^2}\ e^2+\frac{2}{(4\pi)^2}\
\lambda^2)
\end{split}
\end{equation}
This proves that the small velocity anisotropy is
irrelevant.

\bibliography{ncr}
\end{document}